\documentclass[useAMS,usenatbib]{aa}
\usepackage{times}
\usepackage{listings}
\usepackage{graphics}
\usepackage{subfigure}
\usepackage{graphicx}
\usepackage{graphicx,xspace}
\usepackage{amssymb}
\usepackage{amsmath}
\usepackage{mathtools}
\usepackage{hyperref}
\usepackage{aas_macros}
\usepackage{lscape}
\usepackage{rotating}
\usepackage{placeins}
\usepackage{natbib}
\usepackage{color}
\usepackage{caption}
\usepackage{float}
\usepackage{multirow}
\usepackage{threeparttable}
\usepackage{xcolor} 

\bibpunct{(}{)}{;}{a}{}{,}

\begin{document}

\title{Intrinsic alignments of galaxies in multiple projections}

\author{M. L. van Heukelum\inst{1}\fnmsep\thanks{\email{m.l.vanheukelum@uu.nl}}
\and D. Neumann\inst{2}
\and M. Garc\'ia Escobar\inst{1}
\and N. E. Chisari\inst{1,2}
\and H. Hoekstra\inst{2}
}

\institute{Institute for Theoretical Physics, Utrecht University, Princetonplein 5, 3584 CC, Utrecht, the Netherlands    
\and
Leiden Observatory, Leiden University, P.O. Box 9513, 2300 RA Leiden, the Netherlands
 }
    
   \date{Received 09-09-2025; accepted 22-11-2025}

\abstract 
{The intrinsic alignments of galaxies can be measured and modelled to gain cosmological information and further improve our understanding of the interactions between galaxies, as well as to mitigate their effects on gravitational weak lensing studies.
Hydrodynamical simulations are often used to constrain priors or calibrate models.
Therefore, obtaining the maximum amount of information possible from these simulations is imperative.
In this work, we  combined the information of shapes projected over two or three axes ($x,y,z$), for intrinsic alignment signals ($w_\mathrm{g+},\ \tilde{\xi}_\mathrm{g+,2}$), showing a consistent gain in signal-to-noise ratio (S/N) for all cases studied using TNG300-1.
The gain in S/N is found to be higher for the addition of the second projection than for the third, and it is also higher for shapes calculated using the reduced inertia tensor, rather than the simple one.
The two shape samples studied, $n_\star>300$ and $\mathrm{log}(M_\star \ h/\mathrm{M_\odot})>10.5$, where the latter has a much higher signal amplitude, show similar gains in S/N when more projections are added. 
We also modelled the correlation functions with the non-linear alignment model for scales greater than 6\,$\mathrm{Mpc}/h$. 
The S/N gains on the non-linear alignment amplitude, $A_{\rm IA}$, and galaxy bias, $b_{\rm g}$, are higher than those seen for the full measurements, indicating potential advantages for future works, particularly on larger scales with an increased uncertainty on the alignment signals.
Using multiple projection axes increases the overall S/N, enabling a more efficient use of numerically expensive hydrodynamical simulations.}


\maketitle
 
\section{Introduction}

 The deflection of light rays due to the presence of matter between a source and the observer, known as weak gravitational lensing, has been established as one of the most powerful probes for exploring the nature of dark matter and dark energy. It can be used to constrain the parameters of the current, widely used cosmological model \citep{albrecht2006report,Kilbinger_2015}. 
In the 2010s, large-area weak lensing surveys (Stage III surveys) such as the Dark Energy Survey\footnote{https://www.darkenergysurvey.org/} \citep{DES}, Kilo-Degree Survey\footnote{https://kids.strw.leidenuniv.nl/} \citep{KiDS}, and Hyper Suprime-Cam\footnote{https://www.naoj.org/Projects/HSC/} \citep{HSC} were developed. 
In recent years, their results \citep{Heymans21,Secco22,Dalal23,Li23,KiDS+DES} have spurred a discussion of potential tensions in early- and late-time Universe cosmology \citep{Tensions}, opening up new questions and paving the way for Stage IV surveys. 
Upcoming surveys include the Legacy Survey of Space and Time\footnote{https://rubinobservatory.org/explore/how-rubin-works/
lsst} on the NSF-DOE \textit{Vera C. Rubin} Observatory \citep{Ivezic19}, the \textit{Euclid} mission\footnote{https://www.euclid-ec.org/public/data/surveys/} \citep{Laureijs11},  and the \textit{Nancy Roman} Space Telescope\footnote{https://roman.gsfc.nasa.gov/} \citep{Roman}, which allow for precision cosmology to be applied.

The main advantage of weak lensing over other probes of cosmic structure is that it is directly sensitive to mass and, thus, less affected by astrophysical uncertainties \citep{Weinberg2013}. 
However, it is susceptible to a number of potential systematic sources of error, which pose challenges for the use of such surveys. 
One of these sources of systematic error is the intrinsic alignments (IA) of galaxies, a term used to refer to the correlations between galaxy shapes and orientations with each other and the underlying density field due to local processes, such as tidal fields \citep{Troxel_2015,Joachimi_2015,Lamman_2024}.
These localised correlations are similar to the effects of cosmic shear and contaminate weak gravitational lensing observations \citep{Krause_2015}.
Therefore, to fully exploit the potential of weak lensing as a tool for precision cosmology, these non-lensing alignments must either be removed or modelled precisely \citep{Joachimi_2010,Zhang_2010,Troxel_2011}. 
For reviews on intrinsic alignments, we refer to \citet{Troxel_2015,Joachimi_2015,Kirk_2015,Kiessling_2015} and, for a practical guide, we refer to \citet{Lamman_2024}.

Apart from their importance as a source of systematic bias in weak lensing studies, there is cosmological information that can be extracted from the IA signal, allowing it to be used to improve our comprehension of the formation and evolution of the large-scale structure of the Universe \citep[e.g.][]{Chisari_2013,Chisari_Dvorkin_Schimdt_2014,Schmidt_2015,Biagetti_2020,Akitsu_2023,Okumura_2023,kurita2023constraints,Xu_2023}. 
Galaxies of different types, colours and in different environments are also sensitive to intrinsic alignments in different ways, likely connected to their evolution through mergers and torques \cite[e.g.][]{Bate20,Bhowmick20} and to baryonic processes \citep[e.g.][]{Velliscig15,Tenneti17,Soussana20}.
These relations and effects are often studied using hydrodynamical simulations such as IllustrisTNG \citep{nelson_2021}, EAGLE \citep{Eagle_sim}, Horizon-AGN \citep{Dubois_HorizonAGN}, and MassiveBlack-II \citep{Khandai_2015}.

The results and insights obtained from hydrodynamical simulations can be used to calibrate analytical models and constrain the priors or their parameters.
Analytical methods include the linear alignment model \citep{Hirata_2004}, extended to incorporate non-linear contributions \citep{Hirata_2007,Bridle_2007}, which is valid on large scales. 
Additionally, the halo model relates the IA signal at small scales to the distribution of matter in dark matter halos \citep{Schneider_2010,Fortuna_2021} and the EFT model describes the IA signal at intermediate scales \citep{Vlah_2020}. 

For the optimal mitigation of the effects of IA on weak lensing surveys, the parameters of the models need to be determined as precisely as possible.
We need to extract as much information as we can from simulations.
The obvious route would be to run larger simulations, increasing the sample sizes and, therefore, decreasing the noise in the measurements.
However, this is very computationally expensive, making other avenues worth investigating.
In recent years, the avenue of finding the optimal estimator has been explored in multiple works.
Along this line, \citet{singh_2024} developed a new multipole-based estimator that increases the precision of parameter fits to IA models by a factor of $\sim2$.
Recently, \citet{lamman2025optimalintrinsicalignmentestimators} suggested an alternative approach, where the line-of-sight integration limit ($\Pi_{\rm{max}}$) is varied with the separation, $r_{\rm{p}}$.
Similarly, \citet{kumwembe2025enhancingmultipletalignmentmeasurements} improved the S/N for multiplet alignments in DESI using imaging.
In this work, we aim to find a complementary method to increase the amount of information that can be extracted from existing simulations, with a relative low computational cost and using the estimators already available.
The main objective is to improve the S/N of intrinsic alignment correlations by considering a combination of the three possible projections over different axes of the simulation box.
While previous works \citep[e.g.][]{Tenneti_2020} have averaged over projections, the non-trivial cross-covariance between these projections has not been considered before, making the gain in S/N that can be obtained from combining multiple projections unknown.
To gain an understanding of how this method complements different estimators, we considered two well-established estimators: $w_\mathrm{g+}$ and the quadrupole, $\tilde{\xi}_\mathrm{g+,2}$ \citep{singh_2024}.

The rest of the work is structured as follows. 
In this study, we use one of the latest hydrodynamical simulations publicly available, the IllustrisTNG project, which will be described in more detail in Sect. \ref{sec:data}, along with the sample selection. 
Section \ref{sect:methodology} discusses the methods adopted to calculate the shapes of galaxies and perform the two-point measurements, as well as the estimation of the covariance matrix through jackknife resampling and the calculation of the S/N values. 
Section \ref{sec:modelling} introduces the non-linear alignment model, used to fit the measurements. In Sect. \ref{sec:results}, we present the results obtained from the applied methods. 
In Sect. \ref{sec:discussion}, we discuss the results and outline potential avenues for future research and Sect. 7 summarises the conclusions drawn in this work.

\section{Data}
\label{sec:data}

We used the IllustrisTNG hydrodynamic simulation. The particle and group data are described in the release papers by \citet{Springel_2017} and \citet{nelson_2021}.

\subsection{IllustrisTNG}\label{sect:IllustrisTNG}
The IllustrisTNG cosmological simulations (hereafter TNG)  are a series of large magnetohydrodynamical simulations of galaxy formation. These simulations are run using the moving-mesh code AREPO \citep{Springel_2010}. A Monte Carlo tracer particle scheme is utilised during the simulation to track the Lagrangian evolution of baryons \citep{Genel_2013}. Each TNG simulation solves for the coupled evolution of dark matter, cosmic gas, luminous stars, and supermassive black holes from an initial redshift of $z = 127$ to $z = 0$.

The TNG simulations are based on a flat $\Lambda$CDM cosmological model, utilising parameters that are consistent with Planck Collaboration results, with a dark energy density of $\Omega_{\Lambda,0} = 0.6911$, total matter density of $\Omega_{\mathrm{m},0} = 0.3089$, baryon density of $\Omega_{\mathrm{b},0} = 0.0486$, amplitude of the matter power spectrum of $\sigma_8 = 0.8159$, and $n_\mathrm{s} = 0.9667$.

These initial parameters determine the distribution of dark matter and gas at high redshift ($z = 127$) within a periodic simulation box of a given volume. The simulations then evolve these conditions forward in time, following the gravitational interactions between dark matter and gas, as well as the hydrodynamics of the gas, using the AREPO code \citep{nelson_2021}. As the simulation progresses, dark matter and gas collapse into dense regions where stars can form. The simulation outputs snapshots at various redshifts, with a total of 100 snapshots. At each snapshot, the friends-of-friends (FoF) and SUBFIND algorithm are used to identify subhalos and, ultimately, galaxies \citep{SUBFIND_Springel_2001}.

The TNG project is made up of three simulations in periodic boxes with side lengths of $35, 75, 205  \mathrm{cMpc}/h$ apiece and at different resolutions. 
In this work, we focus on TNG300 at the highest resolution (TNG300-1), which has $2500^3$ dark matter particles of a mass of $m_{\mathrm{DM}}=5.9\times 10^7 \mathrm{M_\odot}$ and a box size of $(205\ \mathrm{cMpc}/h)^3$. 
In particular, we examined the snapshot at redshift zero. In the simulation, we looked at two distinct mass cuts for the shape sample, described in Sect. \ref{sec:sampleselection}.

\subsection{Sample Selection}
\label{sec:sampleselection}
To measure the cross-correlation between the galaxy positions and their shapes, we defined two galaxy shape samples and one galaxy position sample. 
In this study, the raw data come from the snapshot data catalogues of TNG300-1 at $z=0$, sourced from publicly available data.\footnote{https://www.tng-project.org/data/}
It is then processed using the equations and methods described in Sect. \ref{sect:methodology}.

While simulated data are not subject to the same biases as observed data, such as the challenge of fitting ellipticities in the presence of pixel noise or imperfect PSF modelling \citep{Samuroff_2021}, they remain susceptible to bias. 
This prompts us to implement quality cuts to ensure the validity of the conclusions that have been drawn. 
A primary source of uncertainty arises from galaxy resolution, which impacts the ellipticity and orientation measurements. 
Galaxies with an insufficient number of particles to provide a meaningful shape measurement affect the ellipticity distribution of the sample. 
Through a series of convergence tests, \citet{Chisari_2015} found that a minimum number of 300 particles per galaxy would be needed to mitigate resolution bias to acceptable levels. 
Consequently, in this study, we followed this practice and applied a cut to the shape sample, selecting only galaxies with more than 300 particles. 

For the position sample, we enforce a quality cut of $n_\star>50$. 
This threshold reflects the minimum number of particles required for a bound substructure to be considered a legitimate galaxy by the SUBFIND algorithm used in the IllustrisTNG simulations \citep{nelson_2021}. 
Galaxies with fewer than 50 particles are not well-resolved and could be considered random fluctuations.
These cuts leave a total of 192,109 galaxies in the shape sample and 371,855 in the position sample. 

In addition to this quality cut, we also consider another shape sample defined by the galaxy stellar mass, $M_\star$, enabling us to investigate whether the possible gain in S/N depends on the signal strength, as high mass galaxies have a stronger alignment signal.
The high-mass shape sample is defined by $M_\star>10^{10.5} \mathrm{M_\odot}/h$, and contains 24,430 galaxies.
The choice in terms of the position sample remains the same.

\section{Methodology}
\label{sect:methodology}

\subsection{Galaxy shapes}
To characterise the three-dimensional (3D) shape of a galaxy, we use both the simple and reduced inertia tensor.
While the simple inertia tensor generally provides a higher IA correlation function signal amplitude, allowing for more insight in the workings of galaxies, the reduced inertia tensor shape gives more weight to particles closer to the centre. 
This leads to rounder estimates of galaxy shapes, which mimic the observations more closely and are, therefore, also worth studying \citep{Tenneti_2015}.
The simple (Eq. (\ref{eq:inertiatensor})) and reduced (Eq. (\ref{eq:reduced inertiatensor})) tensors are given by
\begin{align}
    I^{k,l}_j &= \frac{1}{M_j} \Sigma^{N_j}_{i=1} m_i r^k_i r^l_i\label{eq:inertiatensor},\\ 
     I^{k,l}_j &= \frac{1}{M_j} \Sigma^{N_j}_{i=1} \frac{m_i r^k_i r^l_i}{r^2}, \label{eq:reduced inertiatensor}
\end{align}
where $k,l$ run from $0$ to $2$ and denote the $x,y,z$ components of a vector, $\mathbf{r}$, and $i$ runs over all $N_j$ particles in galaxy, $j$.
The vector, $\mathbf{r}$, refers to the position of each particle ($\mathbf{x}_i$) relative to the centre of mass ($\mathbf{x}_j$), given by $\mathbf{r}_i = \mathbf{x}_i - \mathbf{x}_j$.
The centre of mass is measured via the mass weighted mean of the particle positions in a galaxy ($\mathbf{x}_{j} = \frac{1}{M_j} \Sigma^{N_j}_{i=1} m_i \mathbf{x}_i$).
The variables $M_j$ and $m_i$ refer to the galaxy and particle masses, respectively.

To determine the two-dimensional (2D) projected shapes of the galaxies, the positions of the particles are projected onto a plane, which can be given by either side of the simulation box. 
In most other studies, the $xy$ plane is chosen arbitrarily, and thus Eq. \ref{eq:inertiatensor} is projected along the $z$-axis by using only the $k, l = 0, 1$ elements, corresponding to $x, y$. 
For Eq. \ref{eq:reduced inertiatensor}, $r^2$ is equal to $r_x^2+r_y^2$, and so becomes the projected distance from the particle to the centre of mass.

In this work, we calculated three different 2D projected shapes by projecting onto the $xy, xz$, and $yz$ planes, hereafter called the $z$, $y$ and $x$ projections, respectively, referring to the line-of-sight axis.
The three projected inertia tensors can be diagonalised to yield the eigenvalues, \( \lambda_a > \lambda_b \), which correspond to the squared values of the lengths of the semi-major and semi-minor axes of the projected ellipsoid and the corresponding eigenvectors, \( \hat{e}_a \) and \( \hat{e}_b \), denote their orientation.
We define the axis ratio of the galaxy as $q = b/a$, where $a$ and $b$ represent the semi-major and semi-minor axis lengths ($a = \sqrt{\lambda_a}$, $b = \sqrt{\lambda_b}$).

The radial and tangential ellipticity components of the shape galaxy relative to the position galaxy is then defined by Eq. \ref{eq:ellipticity}. The components of the ellipticity are given by \citet{Mandelbaum_2006}:
\begin{equation}\label{eq:ellipticity}
    (e_+,e_\times) = \frac{1-q^2}{1+q^2}[\cos(2\phi),\sin(2\phi)],
\end{equation}
where $q$ is the axis ratio and $\phi$ is the orientation angle of the semi-major axis (i.e. the angle between the semi-major axis of the galaxy and the separation vector between a given position and shape galaxy pair). 
For this definition, $e_+>0$ indicates radial alignment, $e_+<0$ tangential alignment and $e_\times$ is the cross component, which generally averages to zero in the universe. 
The total ellipticity of a galaxy is then calculated as $e = \sqrt{e_+^2+e_\times^2}$.
We note that we have chosen the convention of using the distortion as the ellipticity, $e=\frac{1-q^2}{1+q^2}$, instead of the other commonly used definition $e=\frac{1-q}{1+q}$.
As we are using $e$ to denote this quantity, we use the term ellipticity throughout the paper.

\subsection{Projected correlations}
\label{sect: wg+ method}
We focussed our analysis on the cross-correlation of galaxy positions and intrinsic ellipticities, $\xi_\mathrm{g+}$. 
To estimate $\xi_\mathrm{g+}$, we adopted a commonly used estimator, the Landy-Szalay (LS) estimator \citep{LandySzalay}. This estimator was originally devised to estimate galaxy clustering ($\xi_{gg}$), but it can be generalised to calculate other correlation functions, including $\xi_\mathrm{g+}$ \citep{Mandelbaum_2006}. 
Alternative estimators, such as the one used in \citet{Joachimi_2011}, can also be employed. 
However, the LS estimator has several advantages, including having a higher signal-to-noise ratio (S/N) compared to other estimators \citep{Singh_Mandelbaum_More_2015}. 

We defined the cross-correlation of galaxy positions and intrinsic ellipticities, $\xi_\mathrm{g+}(r_\mathrm{p}, \Pi)$, as a function of the projected separation, $r_\mathrm{p}$, and along the line of sight, $\Pi$ \citep{Mandelbaum2011}:
\begin{equation}\label{eq:corr_g+}
    \xi_\mathrm{g+}(r_\mathrm{p}, \Pi)= \frac{S_+D-S_+R_D}{R_S R_D},\end{equation}
\noindent where $S_+$ refers to the $+$ component of our galaxy shape sample and $D$ is the position sample. Additionally, we define $R$ as a set of randomly distributed galaxies with no correlations from large-scale clustering. Here, $S_+D$ refers to the sum over all position-shape pairs of simulated galaxies and is given by
\begin{equation}
    S_+D = \sum_{r_\mathrm{p},\Pi} \frac{e_{+}(j|i)}{2\mathcal{R}},\end{equation}
where $i,j$ refer to galaxies in the shape or position sample, respectively.
In this equation, $\mathcal{R}$ refers to the responsivity factor. 
This variable is a measure of the response of the ellipticity to a small shear and is given by $\mathcal{R} = 1 - \langle e^2 \rangle/2$, where $\langle e^2\rangle$ is the mean squared ellipticity of the galaxy sample \citep{Bernstein_Jarvis}. 
$S_+R_D$ refers to the sum over the ellipticities of the galaxies around the random points, which we assumed to be zero, following \citet{Tenneti_2015}. 
The quantities $R_S$ and $R_D$ represent randomly distributed points in the shape sample and density sample, respectively. 
The denominator, $R_S R_D$, corresponds to the expected number of pairs of random galaxies with separation and $r_\mathrm{p}$ and $\Pi$ for a uniform random distribution. 
This quantity is calculated analytically following the approach outlined by \citet{singh_2024}.

From these 3D measurements, we can straightforwardly obtain the 2D projected correlation, $w_\mathrm{g+}$, which is one of the primary targets of this study. 
While the 3D information of galaxy shapes can provide more insight, comparing such data with observations is challenging as all galaxy shapes are projected on the sky and we have a fixed line of sight determined by our position in space. 
Projected correlation functions are more straightforward to observe and model, which makes it more useful to facilitate a direct comparison between simulation results and observational data. 

The projected correlation function between galaxy positions and shapes, $w_\mathrm{g+}$, is given by integrating along the line of sight, 
 \begin{equation}\label{eq:wgplus}
     w_\mathrm{g+} (r_\mathrm{p}) = \int_{-\Pi_{\text{max}}}^{\Pi_{\text{max}}} \xi_\mathrm{g+}(r_\mathrm{p}, \Pi) \,\text{d}\Pi,
 \end{equation}
where $\Pi_{\text{max}}$ is an integration limit, which can be set by the simulation volume, or chosen separately. 
In this work, we take $\Pi_\text{max} = 20\mathrm{Mpc}/h$, to reduce the noise.
The impact of using half the box size ($102.5\mathrm{Mpc}/h$) as $\Pi_{\mathrm{max}}$ on the results is discussed in Appendix \ref{app:pimax}. 
The measurements are performed in 15 radial bins from $0.1 \mathrm{Mpc}/h \leq r_\mathrm{p} \leq 40 \mathrm{Mpc}/h$; determined by the number of jackknife regions (see Sect. \ref{sect:cov}) and the box size, which limits the largest scales we can probe without running out of galaxy pairs. Thus, we obtained three different projected cross-correlation functions, $w_\mathrm{g+}^x$, $w_\mathrm{g+}^y$, and $w_\mathrm{g+}^z$, by projecting along the three axes of the simulation box.

\subsection{Multipoles}

We also measure the multipole expansion of the correlation function, which has emerged as a new and improved estimator for the intrinsic alignment of galaxies. 
\citet{Singh_Mandelbaum_2016} highlighted the potential of using multipole expansion due to the well-defined scaling of the IA signal in the $(r_\mathrm{p}, \Pi)$ plane. 
In such instances, the correlation function can be expressed in terms of their multipole moments according to the following expression \citep{singh_2024,Okumura_2023}:
 \begin{equation}
     \tilde{\xi}_{ab}^{\ell,s_{ab}} (r) = \frac{2\ell + 1}{2}\frac{(\ell - s_{ab})!}{(\ell + s_{ab})!}\int \text{d} \mu_{r}\,L^{\ell,s_{ab}}(\mu_r)\,\xi_{ab}(r, \mu_{r}) 
     \label{eq:multipole_definition}
 ,\end{equation}
  \begin{equation}
    \mathrm{with}\quad \mu_{r} = \frac{\Pi}{r}.
 \end{equation}
Here, $\mu_{r}$ is the cosine of the angle between line-of-sight direction and the separation vector $\bf{r}$. 
$L^{\ell,s_{ab}}$ refers to the $\ell$-th order associated Legendre polynomial, while $s_{ab}$ represents the spin parameter. 
Note that \citep{singh_2024} use a wedged multipole, setting the integrand in Eq. (\ref{eq:multipole_definition}) to zero for $r_\mathrm{p}<2\mathrm{Mpc}/h$. 
We find that for $r>r_\mathrm{p}$ the wedged multipole quickly converges to the full expression, leading to virtually no difference in our results since we restrict the modelling analysis to scales $r>6\mathrm{Mpc}/h$.

In this study, similarly to the case of projected correlations, we focus on the cross-correlation between galaxy positions and shapes, $\tilde{\xi}_\mathrm{g+}$. Galaxy shapes are considered spin-two objects due to their symmetry under rotation by 180 degrees, which implies that $s_{ab} = 2$ for $\tilde{\xi}_\mathrm{g+}^{\ell,s_{ab}}$ \citep{singh_2024}. 
Because of symmetry, the quadrupole moment is the lowest non-zero multipole moment, which leads to $\ell = 2$. 
Therefore, the expression used to calculate the cross-correlation function in terms of the multipole moments is given by $\tilde{\xi}_\mathrm{g+}^{2,2} (r)$, hereafter abbreviated to $\tilde{\xi}_\mathrm{g+,2}$,   \begin{equation}
     \tilde{\xi}_\mathrm{g+}^{2,2} (r) = \frac{5}{48}\int \text{d} \mu_{r}\,L^{2,2}(\mu_r)\,\xi_\mathrm{g+}(r, \mu_{r}) .
 \end{equation}
This approach is a relatively new avenue for estimating intrinsic alignments. \citet{kurita2023constraints} measured the 3D IA power spectrum from the spectroscopic BOSS galaxy sample for the first time by decomposing the spin-2 IA field into multipole moments. 
They showed that this approach leads to higher S/N and tighter constraints. 
Recent work by \citet{singh_2024} has also demonstrated the potential of this approach, showing that using a multipole-based estimator can lead to a $\sim 2.3$ increase in statistical precision of alignment amplitude. 
This improvement is equivalent to having a survey area that is four times larger. 
Motivated by these promising findings, this work aims to investigate the improvement in the IA signal when employing the multipole-based estimator (hereafter referred to as quadrupole) in the context of hydrodynamic simulations. 
Similar to the case of $w_\mathrm{g+}$, we calculate three different projections of the quadrupole, $\tilde{\xi}_\mathrm{g+,2}^x$, $\tilde{\xi}_\mathrm{g+,2}^y$, and $\tilde{\xi}_\mathrm{g+,2}^z$, by projecting along the three axes of the simulation box.

For the measurements of both $w_\mathrm{g+}$ and $\tilde{\xi}_\mathrm{g+,2}$ the code {\sc{MeasureIA}}\xspace\footnote{https://github.com/MarloesvL/measure\_IA} was used.
{\sc{MeasureIA}}\xspace is publicly available and has been validated against {\sc{halotools}}\xspace\footnote{ https://github.com/astropy/halotools} \citep{halotools} and {\sc{TreeCorr}}\xspace\footnote{https://github.com/rmjarvis/TreeCorr} \citep{treecorr} for $w_\mathrm{g+}$ and $w_\mathrm{gg}$ measurements, including the covariance.

\subsection{Covariance matrices}
\label{sect:cov}
The covariance matrices for our measurements are estimated using the jackknife method \citep{Quenouille_1956,tukey_1958,Shao}. 
We divide the simulation box into $N_\mathrm{sub}$ cubes of equal volumes and compute the alignment statistics by omitting one of the $N_\mathrm{sub}$ sub-volumes at a time. 
This process leads to $N_\mathrm{sub}$ jackknife measurements. 
The covariance matrix for $N$ jackknife resamplings is then estimated using the following expression \citep{Norberg_2009}:
\begin{equation}\label{Cjk: simple}
    C_{ij} = \frac{N_\mathrm{sub}-1}{N_\mathrm{sub}} \sum_{k=1}^N (\psi_i^k - \Bar{\psi}_i)(\psi_j^k - \Bar{\psi}_j),
\end{equation}
where $\psi_i$ is the value of $\psi$ for the $i$-th $r$ or $r_\mathrm{p}$ bin of a given statistic of interest, $\psi \in (\tilde{\xi}_\mathrm{g+,2}, w_\mathrm{g+})$, for the $k$-th jackknife resampling. 
The mean value of the $i$-th measurement over all jackknife resamplings, $\Bar{\psi}_i$ , is given by
\begin{equation}
    \label{eq: Jackknife_bar_estimate}
    \Bar{\psi}_i = \sum_{k=1}^{N_{\mathrm{sub}}} \frac{\psi_i^k}{N_\mathrm{sub}} .
\end{equation}
According to \citet[][Appendix D]{Hirata_2004}, $N_\mathrm{sub}$ needs to be larger than $N_\mathrm{bin}^{3/2}$, with $N_\mathrm{bin}$ the number of $r_\mathrm{p}$ bins, and the length of their box needs to be larger than the largest scales measured.
Therefore, for this work, $N_\mathrm{sub} = 5^3 = 125$ sub-volumes are used, and the effect of using $N_\mathrm{sub} = 64$ is explored in Appendix \ref{app:jk}. 

It is important to note that the jackknife error estimation is limited by the size of the simulation box. 
As found by \citet{Norberg_2009}, even though the jackknife covariance estimator is accurate on large scales, it tends to overestimate the covariances on small scales. 
While this effect depends on the size of the $N_\mathrm{sub}$ sub-volumes the data is split into, it is possible that the error bars will be overestimated, especially on smaller scales. 
Unfortunately other estimators also present limitations, and addressing these effects is beyond the scope of this work.
However, the validity of our assumptions has been tested on TNG100-1, which has a smaller box size and higher resolution than TNG300-1.
The results of these tests can be found in Appendix \ref{app:TNG100}.

We can rewrite Eq. (\ref{Cjk: simple}) to include the three different projections of our alignment statistics. The data vector, $\boldsymbol{\psi}$, is redefined as a concatenation of the three possible projections,
\begin{equation}\label{eq:combvector}
    \boldsymbol{\psi} = (\boldsymbol{\psi}^{x}, \boldsymbol{\psi}^{y}, \boldsymbol{\psi}^{z}),
\end{equation}
and we define the new covariance matrix, $C_{ij}^{ab}$, as
\begin{equation}\label{Cjk: combined}
    C_{ij}^{ab} = \frac{N_\mathrm{sub}-1}{N_\mathrm{sub}} \sum_{k=1}^{N_\mathrm{sub}} (\psi_i^{a,k} - \Bar{\psi}_i^a)(\psi_j^{b,k} - \Bar{\psi}_j^b),
\end{equation}
where $a,b \in (x,y,z)$, with nine possible combinations of $a$ and $b$. 
Thus, the full covariance matrix of all three projections combined, $C_{ij}^{xyz}$, is a block matrix, with block elements, $ C_{ij}^{ab}$ 
can be represented as
\begin{equation}
    C_{ij}^{xyz}  =
\begin{bmatrix}
C_{ij}^{xx} & C_{ij}^{xy} & C_{ij}^{xz}\\
C_{ij}^{yx} & C_{ij}^{yy} & C_{ij}^{yz}\\
C_{ij}^{zx} & C_{ij}^{zy} & C_{ij}^{zz}.
\end{bmatrix}.
\end{equation}
This covariance matrix, which incorporates the covariances between different projections, will be referred to as the `combined covariance matrix' in subsequent sections. 
The dimensions of this combined covariance matrix are $3N_\text{bins} \times 3N_\text{bins}$, where each block $C_{ij}^{ab}$ is an $N_\text{bins} \times N_\text{bins}$ matrix. 
In this work, since $N_\text{bins} = 15$, the combined covariance matrix has dimensions of $45  \times 45$, accounting for the 15 radial bins in each of the three projections ($x, y$ and $z$).
When considering the combinations of two projections alone, the relevant four $C_{ij}^{ab}$ blocks can be combined to form the $30\times30$ covariance matrix.

\subsection{Signal-to-noise ratio}
\label{sect:SNR method}
The S/N is generally calculated using the following expression,
\begin{equation}\label{eq:SNR general}
    \left(\frac{\mathrm{S}}{\mathrm{N}}\right)^2 = \sum_{ij}\psi_i C_{ij}^{-1} \psi_j,
\end{equation}
\noindent where $\psi$ represents the $w_\mathrm{g+}$ or $\tilde{\xi}_\mathrm{g+,2}$ data vector, and $C_{ij}^{-1}$ is the inverse of the covariance matrix.
However, when using the jackknife method to estimate the covariance matrix, only a limited resolution of 
\begin{equation}
    \Delta C_{ij} \simeq \sqrt{\frac{2}{N_{\mathrm{sub}}}}
\end{equation}
can be attained, with $N_{\mathrm{sub}}$ the number of jackknife regions, as explained by \citet{Gaztanaga_2005}.
Following the procedure from \citet{ES01,Gaztanaga_2005}, we therefore introduce a resolution condition, imposed on the covariance, which becomes more influential when the projections are combined.
The S/N can then be calculated as follows.
First, the data and the covariance need to be normalised according to
\begin{align}
    \hat{\psi_{i}} = \frac{\psi_{i}}{\sigma_i} \quad \mathrm{and} \quad
    \hat{C}_{ij} = \frac{C_{ij}}{\sigma_i \sigma_j},\label{eq:cov norm}
\end{align}
where $\sigma_i\coloneqq\sqrt{C_{ii}}$ is the variance.
After the normalisation, we perform a Singular Value Decomposition (SVD) of the covariance matrix,
\begin{equation}
    \hat{C}_{ij} = \sum_{kl}(U_{ik})^{\dagger} D_{kl} V_{lj},
\end{equation}
where $D_{kl}$ is, per definition, a diagonal matrix with elements $\lambda_i^2$. 
Note that other  literature commonly refers to $V$ as hermitian instead of $U$. 
In the case of a positive-definite symmetric matrix, such as a covariance, this is irrelevant because $U=V$.
The uncertainty on $\Delta C_{ij}$ can be translated into a condition on the $\boldsymbol{D}$ eigenvalues, given by
\begin{equation}
    \lambda_i^2 > \sqrt{\frac{2}{N_{\mathrm{sub}}}},
    \label{eq:cov res}
\end{equation}
which we motivate more in-depth in Appendix \ref{app: Derivation of the resolution cut}. We conduct the rest of the analysis in the basis where the covariance is diagonalised. Therefore, the new data vector elements become
\begin{equation}
    \tilde{\psi_i} = \sum_j U_{ji} \hat{\psi_j}.
    \label{eq: vector rebase}
\end{equation}
When using the inverse of an estimated covariance, as we need for the S/N (Eq. (\ref{eq:SNR general})), we apply the Hartlap factor to correct for the bias introduced by the inversion \citep{Hartlap_2006}.
This factor is given by
\begin{equation}
    f_H = \frac{N_\mathrm{{sub}}-p-2}{N_\mathrm{{sub}}-1},
\end{equation}
where $p$ is the number of elements in the data vector.
It is only valid when $p<N_\mathrm{{sub}}-2$, which is true in our case.
As we are effectively reducing the number of elements in the data vector by imposing the resolution condition, we are using the number of elements left after the condition is included as our $p$, which is consistent with the work of \citet{Philcox_2021}.
As the data vector length changes with a factor of $2$ or $3$, depending on the number of projections combined, but $N_\mathrm{{sub}}$ remains the same, the Hartlap factor will be different according to the number of projections.

To calculate the S/N, we use only the elements $i$ of the new data vector for which Eq. \ref{eq:cov res} is satisfied ($p$), summing over the squared values of the $(\mathrm{S/N})_i$, which is the S/N for each individual element. The total S/N then becomes
\begin{align}
    \label{eq: SNR_definition}
    \frac{\mathrm{S}}{\mathrm{N}} = \sqrt{f_H\sum_i \left(\frac{\mathrm{S}}{\mathrm{N}} \right)_i^2}=\sqrt{f_H\sum_i \left| \frac{\tilde{\psi_i}}{\lambda_i} \right| ^2} .
\end{align}
In Sect. \ref{sect:results_SNR}, this version of the S/N is used to compare between cuts, using one or more projections.

\section{Modelling}
\label{sec:modelling}
We fit the correlation functions using the linear alignment (LA) model \citep{Catelan_2001}. This model is based on the premise that there exists a linear relation between the intrinsic shapes, or ellipticities, of galaxies and the tidal field at the time of galaxy formation. Such linear relation, as expressed by \citet{Singh_Mandelbaum_More_2015}, is:
\begin{equation}
    \gamma^I = (\gamma_+^I,\gamma_\times ^I) = - \frac{C_1}{4\pi G}(\partial_x^2-\partial_y^2,\partial_x\partial_y)\phi_p , 
\end{equation}
\noindent where $\phi_p$ is the primordial gravitational potential, $C_1$ is a constant that determines the strength of the alignment, and $G$ is Newton’s gravitational constant. 

Starting from this equation, \citet{Hirata_2004} derived the power spectrum for the two-point position-shape correlation functions, linking them to the linear matter power spectrum, $P_\delta^{\mathrm{lin}}$. Focussing only on the cross-correlation, the cross-power spectrum between the galaxy density field and the galaxy intrinsic shear along the line connecting a pair of galaxies is given by \citet{Singh_Mandelbaum_More_2015}:
\begin{equation}\label{eq:powerspectrum}
    P_\mathrm{g+}(\textbf{k},z) = A_{\mathrm{IA}}\,b_\mathrm{g} \frac{C_1\,\rho_{\mathrm{crit}}\,\Omega_\mathrm{m}}{D(z)}\frac{k_x^2-k_y^2}{k^2}P_\delta^{\mathrm{lin}}(\textbf{k},z).
\end{equation}
Here, $b_\mathrm{g}$ represents the linear galaxy bias, which relates the galaxy density field to the underlying matter density field. In line with other studies, we assume that $b_\mathrm{g}$ is scale-independent \citep{Joachimi_2011,Chisari_2013}. $D(z)$ is the growth function, normalised to unity at $z = 0$, to account for the fact that the LA model assumes alignments are established at the time of galaxy formation. Following the conventions of \citet{Joachimi_2011}, we set $C_1\rho_{\mathrm{crit}} = 0.0134$. As defined by \citet{Singh_Mandelbaum_More_2015}, we use a free constant $A_\mathrm{IA}$ to measure the amplitude of IA.
In practice, we use Eq. (3.4) from \citet{Blazek_2011} and Eq. (9) from \citet{singh_2024} to calculate $w_\mathrm{g+}$ and $\xi_\mathrm{g+}^{2,2}$ predictions, respectively, without a Limber approximation.

While the LA model is reasonable at large scales, its reliability diminishes on small scales. For this reason, \citet{Bridle_2007} propose, following \citet{Hirata_2004}, to replace $P_\delta^{\mathrm{lin}}$ with the non-linear power spectrum $P_\delta^{\mathrm{nl}}$, resulting in what is known as the non-linear alignment (NLA) model. This provides a more realistic modelling of intrinsic alignments on small scales, where non-linear effects become significant. In this work, we use the NLA model to calculate the predicted values for $w_\mathrm{g+}$ and $\tilde{\xi}_\mathrm{g+,2}$. The non-linear matter power spectrum will be calculated through the Core Cosmology Library (CCL)\footnote{https://github.com/LSSTDESC/CCL} \citep{Chisari_2019}, which uses the software CAMB code to calculate the values \citep{Lewis_2002}. 

In our analysis, we fit for a single parameter in the NLA model, which is the product of the galaxy bias, $b_\mathrm{g}$ and the IA amplitude, $A_{\rm IA}$, henceforth designated as NLA amplitude. For each dataset, we perform the fitting procedure in two different ways. First, we fit each of the three individual projections ($x$, $y$ and $z$) separately, using the corresponding individual covariance matrix. The resulting best-fit parameter from this approach is referred to as $b_\mathrm{g}A_{\mathrm{IA}} (1)$. Secondly, we fit a combination of all three projections simultaneously, where the obtained best-fit parameter is denoted as $b_\mathrm{g}A_{\mathrm{IA}} (3)$. 

The best-fit parameter for $b_\mathrm{g}A_{\mathrm{IA}} (1)$ and $b_\mathrm{g}A_{\mathrm{IA}} (3)$, along with their uncertainties, were determined by minimising the chi-squared statistic. Similar to the S/N considerations from Sect. (\ref{sect:results_SNR}), we transformed data and model vector according to Eq. (\ref{eq: vector rebase}) and apply the resolution condition Eq. (\ref{eq:cov res}). The transformed chi-squared then becomes
\begin{equation}
    \label{eq:chi2 transformed}
    \tilde{\chi} ^2 = \sum_i\left(\frac{(\tilde{\psi}_\mathrm{g+})_i-(\tilde{\psi}_\mathrm{g+,\text{model}})_i}{\lambda_i}\right)^2,
\end{equation}
\noindent where $\lambda_i$ are the Eigenvalues of the decomposed covariance matrix $C_{ij}$. 
In the case of the individual fitting, $C_{ij}$ refers to the corresponding individual covariance matrix, while for the combined fit of three projections, $C_{ij}$ refers to the combined covariance matrix, as described in Eq. (\ref{Cjk: combined}).

The NLA model is known to break down and underestimate the alignment signal on small scales. 
To mitigate this, we follow the approach adopted by many studies and restrict the fitting range for the NLA model to intermediate and larger scale (see \citealt{singh_2024}, \citealt{Tenneti_2015}, among others). 
We fit the model only over the region between 6 Mpc$/h$ $< r_\mathrm{p} <$ 40 Mpc$/h$.

\section{Results}
\label{sec:results}

\subsection{Projected alignment estimators}
\begin{figure} 
    \centering
    \includegraphics[width=0.5\textwidth]{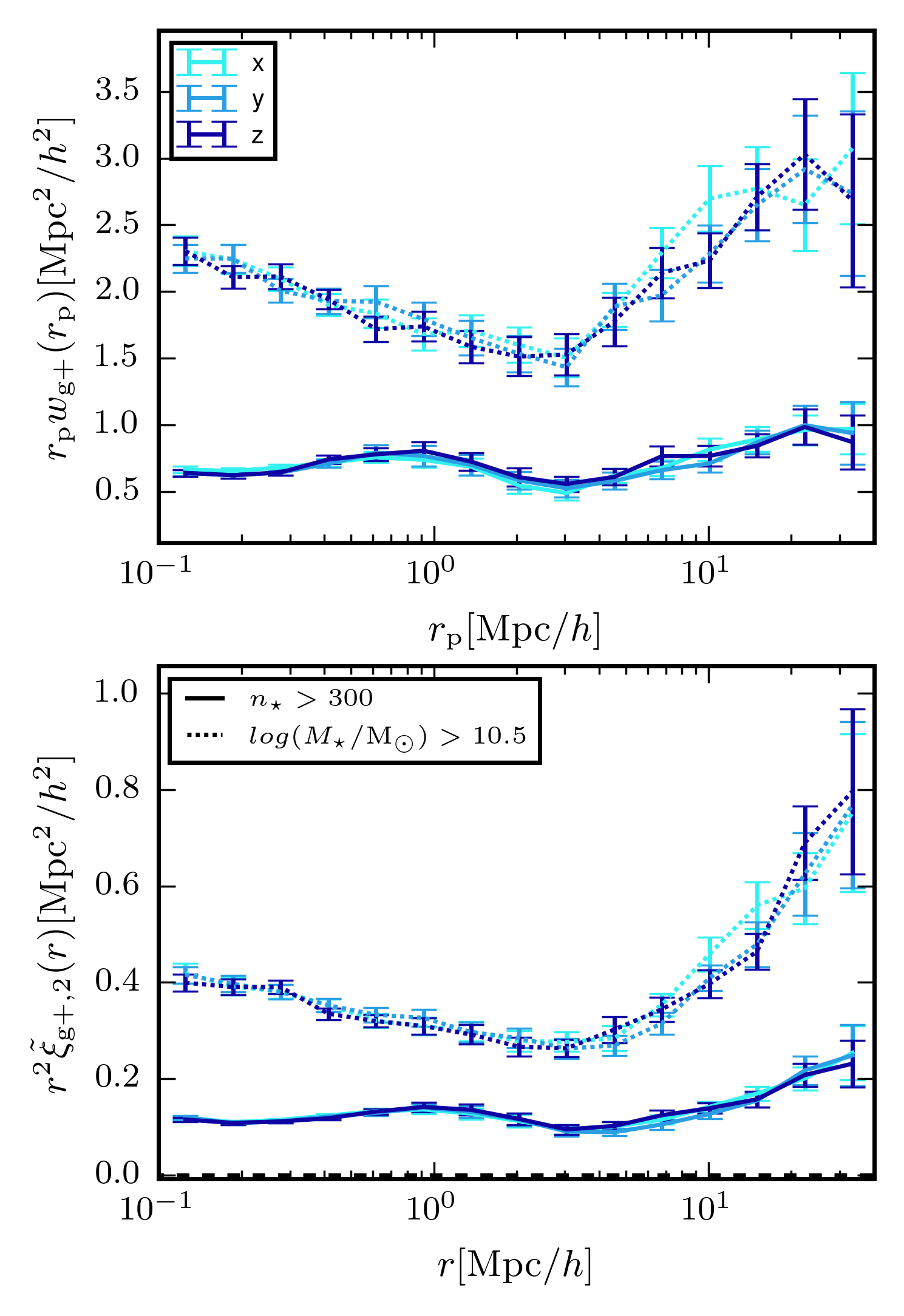}
    \caption{Correlation functions, $r_\mathrm{p} w_\mathrm{g+}$ (top) and $r^2 \tilde{\xi}_\mathrm{g+,2}$ (bottom), in TNG300 for two shape samples: $n_\star>300$ (continuous lines) and $\mathrm{log}(M_\star \ h/\mathrm{M_\odot})>10.5$ (dashed lines).
    The shapes are projected over the $x$ (light blue), $y$ (medium blue), and $z$ (dark blue) axes and measured using the simple inertia tensor.
    }
    \label{Fig:IA TNG300}
\end{figure}

We begin by presenting the projected density-shape correlation function in Fig. \ref{Fig:IA TNG300}, $w_\mathrm{g+}$ (top panel), and the quadrupole, $\tilde{\xi}_\mathrm{g+,2}$ (bottom panel) for three different projections along the $x$ (light blue), $y$ (medium blue) and $z$ (dark blue) axes.
In this figure, $w_\mathrm{g+}$ is multiplied by $r_\mathrm{p}$ and $\tilde{\xi}_\mathrm{g+,2}$ by $r^2$ to negate the main $r_{(\mathrm{p})}$-dependence, making the comparison between different cuts and projections more clearly visible. 
The error bars have been calculated using the jackknife resampling technique described in Sect. \ref{sect:cov}.
Both statistics are shown for two different shape samples, measured using the simple inertia tensor: galaxies with more than 300 stellar particles ($n_\star>300$, continuous lines) and galaxies with $\mathrm{log}(M_\star [\mathrm{M}_\odot/h])>10.5$ (dashed lines).

Figure \ref{Fig:IA TNG300} shows that the three different projections along the $x$, $y$ and $z$ axes are consistent with each other, although at large scales ($r_{(p)}\gtrsim8\,\mathrm{Mpc}/h$) the data points become noisy due to lack of galaxy pairs and sometimes only the error bars overlap, indicating the errors may be underestimated.
The agreement between the signals confirms the expected behaviour that there should be no significant difference between the various line-of-sight projections and that the common choice of projecting onto the $xy$ plane is arbitrary. 
Furthermore, it can be observed that within the same simulation, the error bars on the quadrupoles are smaller compared to those on $w_\mathrm{g+}$, and even though the signal amplitude is lower too, the mean S/N is higher as can be seen in Sect. \ref{sect:results_SNR}. 
This highlights the benefits of using the quadrupole as potential alternative estimators for the density-shape correlation, as suggested by \citet{singh_2024}.

Our choice of $\Pi_{\mathrm{max}}=20\,\mathrm{Mpc}/h$ for $w_\mathrm{g+}$ means that the shapes of the correlation functions are very similar between $w_\mathrm{g+}$ and the quadrupole up to separations of $20 \mathrm{Mpc}/h$ (provided we take out the main $r_{(\mathrm{p})}$-dependence).
Choosing a larger value for $\Pi_{\mathrm{max}}$, adds more noise at large scales to $w_\mathrm{g+}$, as can be seen in Fig. \ref{Fig:IA TNG300 pimax}.

Comparing the mass cuts, we observe that the high mass sample indeed produces a much larger signal amplitude for both $w_\mathrm{g+}$ and $\tilde{\xi}_\mathrm{g+,2}$.
We also see that the error bars for the high mass cut are larger, due to the smaller sample size.
As shown in Sect. \ref{sect:results_SNR}, this stronger signal does not lead to a higher S/N, because both the signal and noise increase by a similar factor.

These signals have also been measured for shapes measured by the reduced inertia tensor.
As this produces rounder shapes, the amplitudes of those signals are much lower.
However, the comparison between the different cuts and projections is very similar to Fig. \ref{Fig:IA TNG300}.
Furthermore, IA correlation functions have been measured in TNG100 for both mass cuts and shape types.
Therein, the correlations are much noisier due to the smaller box size, which also leads to a decreased signal strength.
Barring that, the signals show the same main conclusions as for TNG300.
The S/N values for TNG100 are shown in Appendix \ref{app:TNG100}.

\subsection{Combined covariance matrix}
\label{sect:cov results}

\begin{figure}
    \centering
    \includegraphics[width=0.5\textwidth]{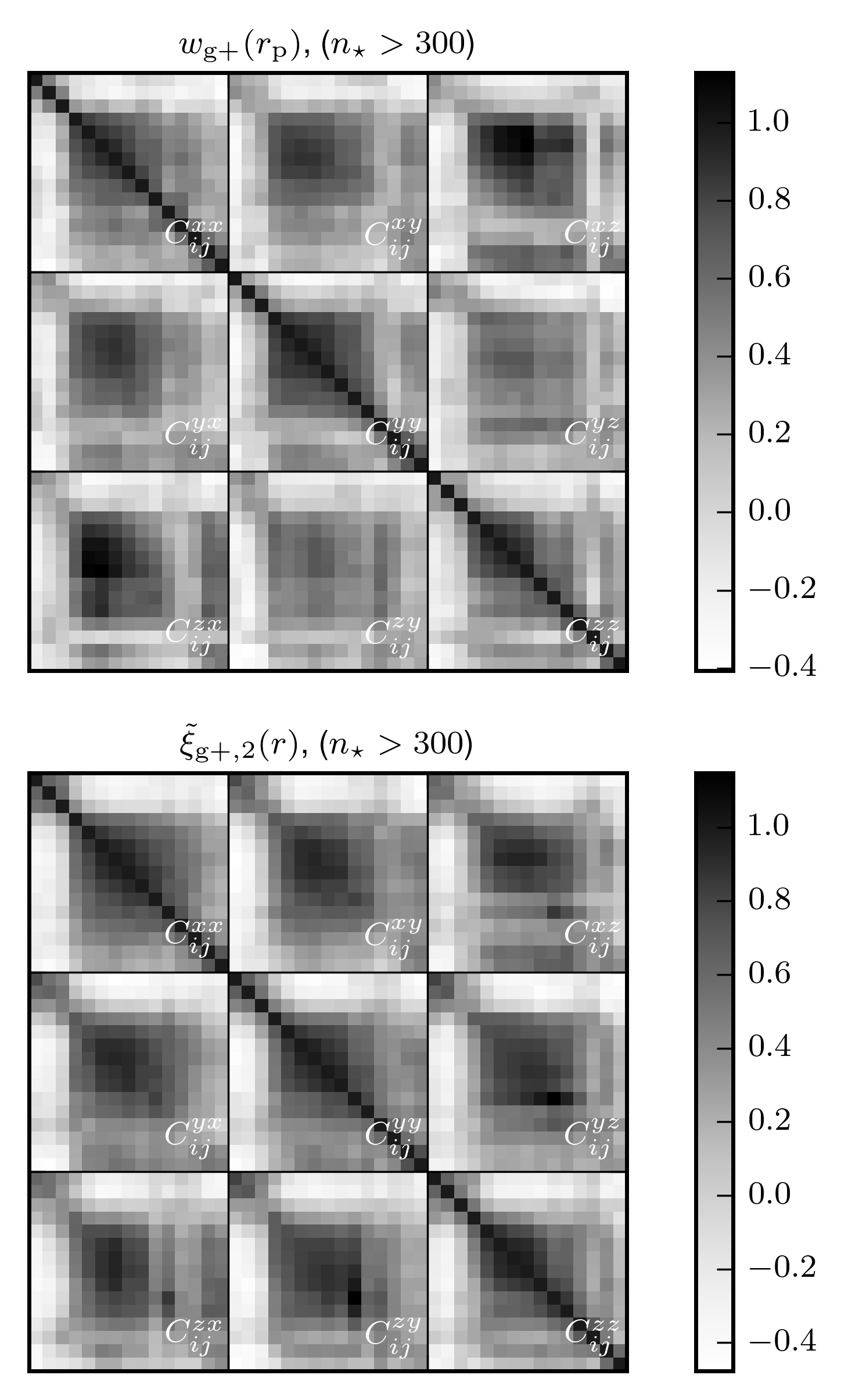}
    \caption{Normalised combined covariance matrices for $w_\mathrm{g+}$ (top panel) and $\tilde{\xi}_\mathrm{g+,2}$ (bottom panel) for the $n_\star>300$ shape sample.
    Darker colours indicate higher values.
    Each block of the combined covariance matrix is labelled according to the definitions in Sect. \ref{sect:cov}.
    }
    \label{Fig:covariancematrix}
\end{figure}

Figure \ref{Fig:covariancematrix} illustrates the normalised combined covariance matrix, as defined in Eq. (\ref{Cjk: combined}) and normalised by Eq. (\ref{eq:cov norm}), for the shape sample defined by $n_\star>300$ for $w_\mathrm{g+}$ (top panel) and $\tilde{\xi}_\mathrm{g+,2}$ (bottom panel). 

For both $w_\mathrm{g+}$ and the quadrupole, the combined covariance matrix exhibits the expected symmetry along the diagonal of the auto-correlation blocks (e.g. $C^{xx}$). 
Comparing the $w_\mathrm{g+}$ and the quadrupole covariance matrices, we see notable differences between the two.
The quadrupole (bottom panel) shows a similar structure in all the auto-correlation and cross-correlation blocks: a high value diagonal; low values for cross-terms with small scales; high values for intermediate scales, $r\gtrsim0.5\,\mathrm{Mpc}/h$, that gradually decrease as the scales increase.
This basic structure is visible in every block, where naturally the central diagonal (auto-correlation blocks) is highest, as it is set to $1$ by definition.
For $w_\mathrm{g+}$ (top panel), we see a less coherent structure arise.
The cross-correlation (off-diagonal) blocks do not show a clear diagonal and while the 'middle' scales have higher values than the cross terms with small or large scales, the normalised values are lower than for the quadrupole; the median value of the ratio between $w_\mathrm{g+}$ and $\tilde{\xi}_\mathrm{g+,2}$ of all elements ($C_{ij}(w_\mathrm{g+})/C_{ij}(\tilde{\xi}_\mathrm{g+,2})$) is $\sim0.84$.
Therefore, it seems that the signals of different projections are less correlated for $w_\mathrm{g+}$ than for the quadrupole.

In Appendix \ref{app:jk}, where the effects of using 64 jackknife regions (instead of 125) are explored, we see that the normalised values of the covariance matrix are generally lower than in Fig. \ref{Fig:covariancematrix}.
This indicates either that choosing 125 jackknife regions overestimates the covariance, or that choosing 64 jackknife regions underestimates the covariance.
We have opted for 125 jackknife regions in this work because these covariance matrices are more stable (see Appendix \ref{app:jk}).
Furthermore, Appendix \ref{app:av cov} explores the effect of averaging the diagonal and off-diagonal covariance blocks, which should be equal, given that there is no preferred line-of-sight direction in a large enough box.
This reduces the noise of the covariance estimate, but erases the influence of the variation in the large scale structure that arises due to the limited box size.
 
\begin{table}
\begin{threeparttable}
\footnotesize
\caption{S/N gain.}
\begin{tabular}{llllll}

\multicolumn{6}{l}{}
\\ \hline\hline
  Stat.   & Shape type & Cut   & 1 vs 3  & 1 vs 2 & 2 vs 3  \\  \hline 
\multirow{4}{*}{$w_\mathrm{g+}$}    & \multirow{2}{*}{simple}    & $n_\star>300$                                  & 1.23   & 1.18             & 1.04           \\ 
                                                                        &                                                                    & $M_\star>10^{10.5}M_\odot/h $         & 1.21  & 1.17             & 1.04          \\  
                                                                        & \multirow{2}{*}{reduced}  & $n_\star>300$                                                                             & 1.27     & 1.19             & 1.06          \\ 
                                                                        &                                                                         & $M_\star>10^{10.5}M_\odot/h $                                                                                 & 1.23 & 1.17             & 1.05          \\  
\multirow{4}{*}{$\tilde{\xi}_\mathrm{g+}$}    & \multirow{2}{*}{simple}    & $n_\star>300$                                & 1.15     & 1.1             & 1.04           \\  
                                                                        &                                                                    & $M_\star>10^{10.5}M_\odot/h $           & 1.15  & 1.11             & 1.03        \\  
                                                                        & \multirow{2}{*}{reduced}  & $n_\star>300$                                                                                & 1.18   & 1.18             & 1.0         \\  
                                                                        &                                                                         & $M_\star>10^{10.5}M_\odot/h $                                                                                 & 1.21  & 1.13             & 1.07         \\  \hline 
\end{tabular}
\begin{tablenotes}
\footnotesize
\item The mean gain in S/N when comparing the use of one versus three; one versus two or two versus three projections.
The gain is shown for $w_\mathrm{g+}$ and $\tilde{\xi}_\mathrm{g+,2}$, for simple and reduced shapes and for both shape samples (mass cuts).
\end{tablenotes}
\label{table:SNR gain}
\end{threeparttable}
\end{table}

\subsection{Signal-to-noise ratio comparison}\label{sect:results_SNR}
We compare the S/Ns of using one or more projections, defined in Sect. \ref{sect:SNR method}. 
A higher S/N indicates a more robust detection of the intrinsic alignment signal against the background noise. 

\begin{figure}
    \centering
    \includegraphics[width=0.5\textwidth]{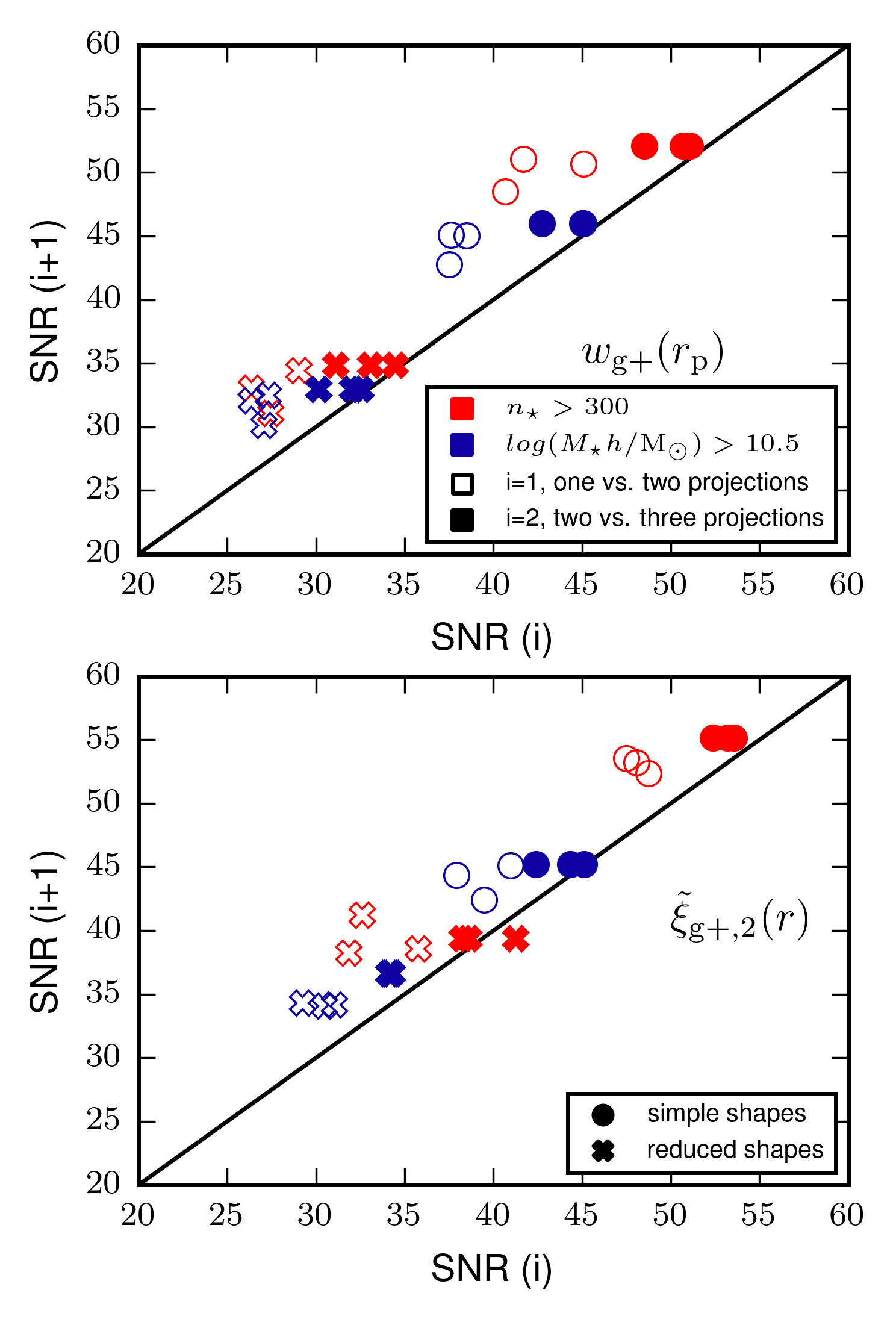}
    \caption{
    S/Ns for the $w_\mathrm{g+}$ (top) and $\tilde{\xi}_\mathrm{g+,2}$ (bottom) of combining one vs two (unfilled) and two vs three (filled) projections.
    The colours of the markers correspond to the different shape samples: $n_\star>300$ is red and $\mathrm{log}(M_\star h/\mathrm{M_\odot})>10.5$ is blue.
    Shapes measured using the simple, reduced inertia tensor are depicted by circles, crosses, respectively.
    The black line denotes the boundary marking where both S/N values are equal.
    }
    \label{Fig:SNR}
\end{figure}

We calculate the S/N for three cases: once for each individual projection $\{x, y, z\}$, S/N (1); three times for a combination of two projections, S/N (2) $\{(x,y), (y,z), (x,z)\}$; and once using the combined covariance matrix and concatenated data vector from all three projections, S/N (3). 
If the signals from the different projections $\{x, y, z\}$ are completely uncorrelated, the maximum improvement in S/N that we can expect by combining the three projections is a factor of $\sqrt3 \approx 1.73$ compared to using any individual projection alone, or $\sqrt2 \approx 1.41$ for two projections. The results are shown in Fig. \ref{Fig:SNR} and Table \ref{table:SNR gain}.

Figure \ref{Fig:SNR} shows the improvement in S/N when one extra projection is added for $w_\mathrm{g+}$ (top panel) and $\tilde{\xi}_\mathrm{g+,2}$ (bottom panel).
In other words, the unfilled markers show the S/N for one projection on the horizontal axis versus two projections on the vertical axis, whereas the filled markers show two versus three projections, respectively.
The colours denote the shape sample that is used: red for $n_\star>300$ and blue for $\mathrm{log}(M_\star \ h/\mathrm{M_\odot})>10.5$.
Furthermore, the circles represent the measurement of the shapes using the simple inertia tensor (Eq. (\ref{eq:inertiatensor})) and the crosses, the reduced inertia tensor (Eq. (\ref{eq:reduced inertiatensor})).
The black line is plotted at S/N(i)=S/N(i+1).
Therefore, an improvement in S/N is shown by the marker residing in the top left half of the plot.
For the one versus two projection markers (unfilled), the combinations $\{x, (x,y)\}$, $\{y, (y,z)\}$ and $\{z, (x,z)\}$ are shown.

For both $w_\mathrm{g+}$ and $\tilde{\xi}_\mathrm{g+,2}$, almost all cases show a clear improvement in S/N when the information of an extra projection is added.
For clarity, Table \ref{table:SNR gain} has been added, which shows the gain in S/N, averaged over the three combinations that are shown individually in Fig. \ref{Fig:SNR} for each choice in shape sample, statistic and shape type.
Together, Fig. \ref{Fig:SNR} and Table \ref{table:SNR gain} reveal that for all our cuts and shape types, we can expect a gain in S/N when combining two or three projections, where adding the third projection to two projections will result in less gain than adding a second projection to one projection (the filled markers are closer to the line than the unfilled ones in Fig. \ref{Fig:SNR}).
The gain in S/N and the S/N values themselves are similar for both mass cuts, despite the high mass sample having a much higher signal amplitude.
Furthermore, the gain in S/N for the reduced shapes is generally higher than the gain in S/N for the simple shapes.
Finally, the gain in S/N is higher for $w_\mathrm{g+}$ than for the quadrupole, although the quadrupole still has a larger S/N in all cases.

One possible reason for these differences in gain is the influence of the covariance resolution criterion in the S/N calculation as described in Sect. \ref{sect:SNR method}.
The number of terms included in the S/N is always higher for $w_\mathrm{g+}$ than for $\tilde{\xi}_\mathrm{g+,2}$ and higher for reduced shapes than for simple shapes.
As each term included into the S/N increases the S/N, more terms added will increase the chances in S/N gain when projections are combined.

\subsection{Model and fit}

\begin{table*}
\centering
\begin{threeparttable}
\footnotesize
\caption{Fitted amplitude.}
\begin{tabular}{llllllll}
\multicolumn{8}{l}{}
\\ \hline\hline
Stat. & Shape type & Cut   & $A_\mathrm{IA}b_\mathrm{g}(x)$  & $A_\mathrm{IA}b_\mathrm{g}(y)$ & $A_\mathrm{IA}b_\mathrm{g}(z)$ & Combined & $\chi^2_\mathrm{red}$ \\  \hline 
\multirow{4}{*}{$w_\mathrm{g+}$}    & \multirow{2}{*}{simple}    & $n_\star>300$                                  & $2.65\pm0.20$  & $2.56\pm0.20$            &  $2.82\pm0.23$ &     $2.6\pm0.15$    & 0.55   \\  
                                                                        &                                                                    & $M_\star>10^{10.5}\mathrm{M_\odot}/h $         & $8.63\pm0.66$  & $7.83\pm0.61$             & $8.4\pm0.51$ &    $8.38\pm0.40$ & 0.75    \\ 
                                                                        & \multirow{2}{*}{reduced}  & $n_\star>300$                                                                             & $0.621\pm0.084$     & $0.496\pm0.083$             & $0.603\pm0.080$ & $0.544\pm0.063$    & 0.83     \\  
                                                                        &                                                                         & $M_\star>10^{10.5}\mathrm{M_\odot}/h $                                                                                 & $3.32\pm0.38$ & $2.76\pm0.37$             & $3.01\pm0.29$ &   $2.98\pm0.23$    & 0.47     \\  
\multirow{4}{*}{$\tilde{\xi}_\mathrm{g+,2}$}    & \multirow{2}{*}{simple}    & $n_\star>300$     & $2.24\pm0.16$& $2.06\pm0.15$& $2.30\pm0.17$ &      $2.22\pm0.13$   & 0.99   \\  
                                                                        &                                                                    & $M_\star>10^{10.5}\mathrm{M_\odot}/h $           & $6.59\pm0.39$&  $6.18\pm0.38$& $6.52\pm0.39$&    $6.43\pm0.27$ & 1.59\\  
                                                                        & \multirow{2}{*}{reduced}  & $n_\star>300$                                                                                & $0.458\pm0.055$& $0.387\pm0.054$& $0.477\pm0.054$&   $0.410\pm0.035$& 1.15\\  
                                                                        &                                                                         & $M_\star>10^{10.5}\mathrm{M_\odot}/h $                                                                                 & $2.47\pm0.24$& $2.27\pm0.22$& $2.25\pm0.20$ &  $2.35\pm0.15$ & 0.85      \\  \hline 
\end{tabular}
\begin{tablenotes}
    \footnotesize
    \item Fitted amplitude for each individual projection as well as for a joint fit considering the full covariance. The amplitude is shown for $w_\mathrm{g+}$ and $\tilde{\xi}_\mathrm{g+,2}$, for simple and reduced shapes and for both shape samples (mass cuts). The shown $\chi^2_\mathrm{red}$ is obtained from the joint fit of the respective three combined projections.
\end{tablenotes}
\label{tab:amplitude_fit_results}
\end{threeparttable}
\end{table*}

In this section we present the results of fitting the NLA model to $w_\mathrm{g+}$ and $\tilde{\xi}_\mathrm{g+,2}$. First, we assess the quality of the fit and then investigate the improvement of S/N when combining projections, analogous to Sect. \ref{sect:results_SNR}.
However, in this section we restrict the scales to $r_{(\mathrm{p})}>6\,\mathrm{Mpc}/h$, only investigating the S/N obtained with common NLA modelling choices.
Therefore, the results shown here serve two purposes, first, to investigate how different scales contribute to the gain in S/N, and second, to examine how the pure gain in S/N propagates into uncertainty of the inferred NLA amplitude.

\begin{figure} 
    \centering
    \includegraphics[width=0.5\textwidth]{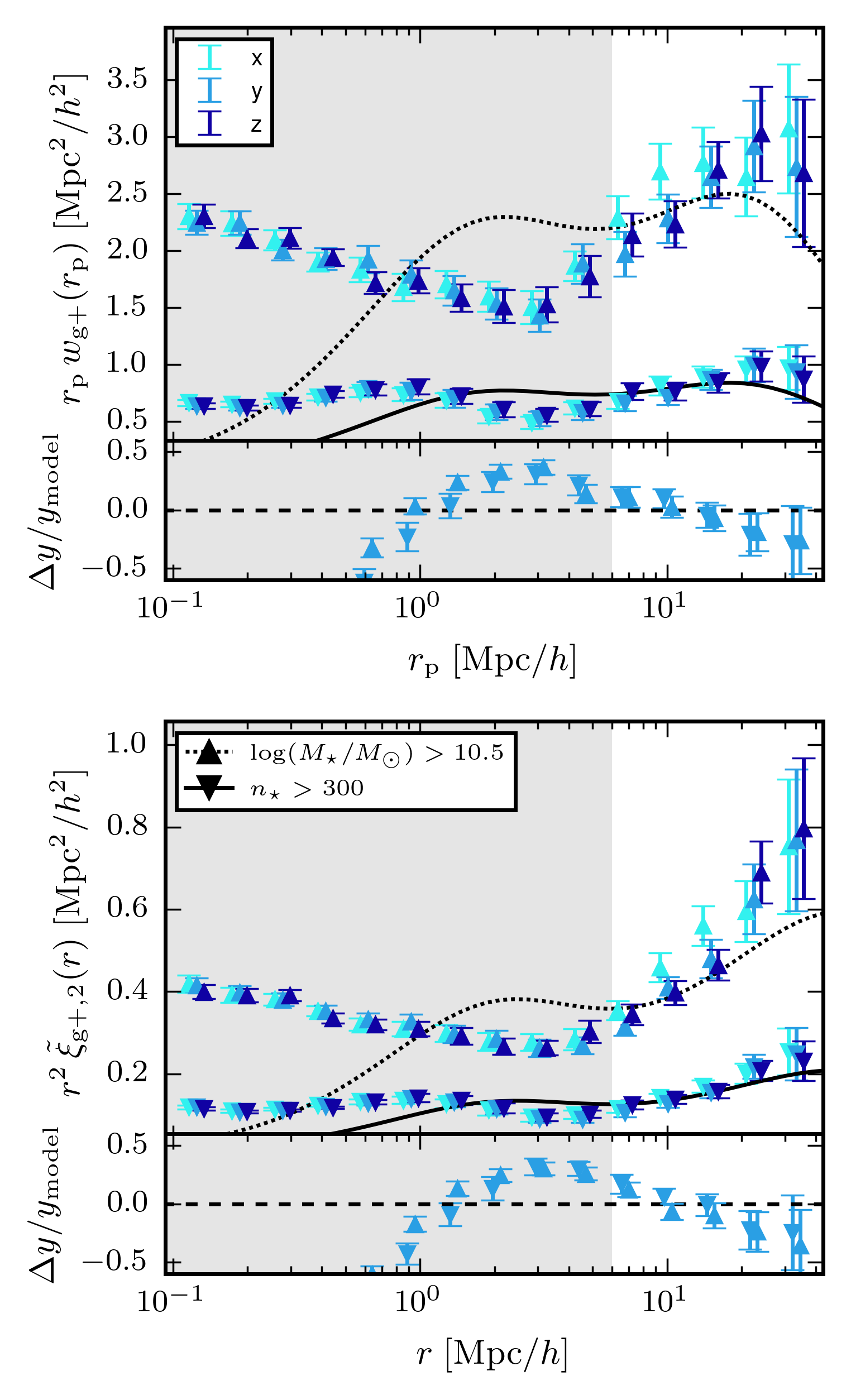}
    \caption{Correlation functions, $r_\mathrm{p} w_\mathrm{g+}$ (top) and $r^2 \tilde{\xi}_\mathrm{g+,2}$ (bottom) in TNG300 for two shape samples: $n_\star>300$ (downward triangle) and $\mathrm{log}(M_\star \ h/\mathrm{M_\odot})>10.5$ (upward triangle). The shapes are projected over the $x$ (light blue), $y$ (medium blue) and $z$ (dark blue) axes and measured using the simple inertia tensor, whereas the joint NLA fit is shown as a solid (dashed) line for the cut in $n_\star$ ($M_\star$). The second panel in both plots shows the fit residual for the $y$-projection. Grey regions indicate excluded regions for the fit. Note: the data points are slightly horizontally displaced for easier readability.
    }
    \label{Fig:IA TNG300 fit}
\end{figure}

\begin{figure}
    \centering
    \includegraphics[width=0.5\textwidth]{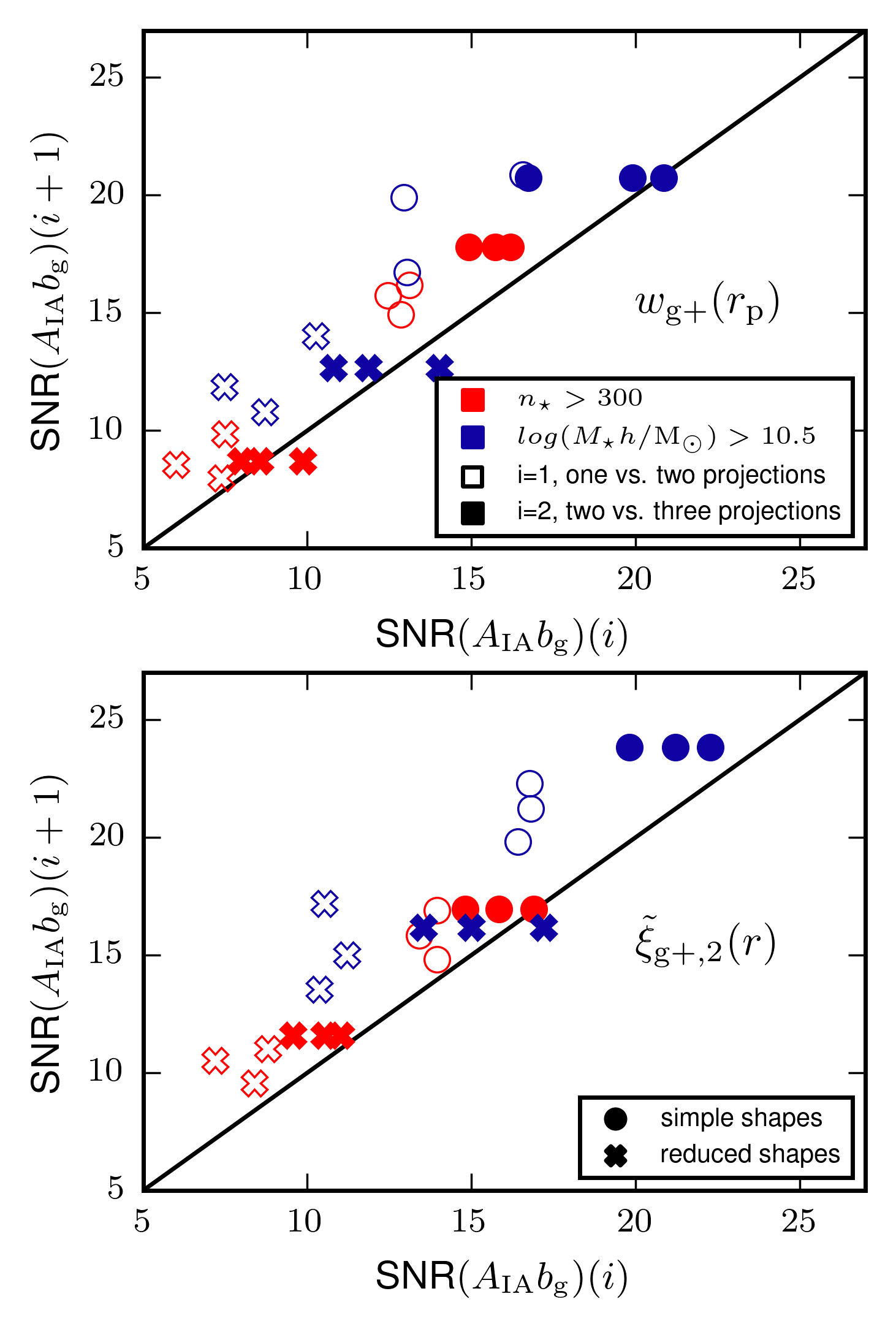}
    \caption{
    Same as Fig. \ref{Fig:SNR}, but for the S/N defined in Eq. (\ref{eq:SNR fitted scaling}) obtained from the non-linear alignment model fit.}
    \label{Fig:SNR from fit}
\end{figure}

Figure \ref{Fig:IA TNG300 fit} shows the combined fit along the x-, y- and z-axis (light, medium and dark blue) for $w_\mathrm{g+}$ (top panel) and $\tilde{\xi}_\mathrm{g+,2}$ (bottom panel) galaxy-shape correlation functions, both for particle number (downward triangle) and mass cut (upward triangle) using the simple inertia tensor definition. 
The NLA model is fitted in the unshaded region and displayed as solid line for the particle number cut and dashed line for the mass cut. 
It must be noted that the individual points in Fig. \ref{Fig:IA TNG300 fit} are highly correlated, not only between different projection directions, but also between $r_\mathrm{(p)}$ bins due to mode mixing. 
However, we can judge the goodness of fit with the reduced $\chi^2_\mathrm{red}$ shown in Table \ref{tab:amplitude_fit_results}.
It ranges from 0.47 to 1.59 and is generally lower for $w_\mathrm{g+}$, possibly indicating slightly over-predicted error bars, but nevertheless an acceptable fit.

The fitted amplitudes for all available measurements are displayed in Table \ref{tab:amplitude_fit_results} as well.
We find a consistent $\sim25\%$ difference in inferred $A_\mathrm{IA}b_\mathrm{g}$ between $w_\mathrm{g+}$ and $\tilde{\xi}_\mathrm{g+,2}$. We believe that this is caused by both methods probing different scales and further discuss this in Appendix \ref{app:fitted amplitude difference}.
Still, the inferred $A_\mathrm{IA}b_\mathrm{g}$ are internally consistent for the individual projection directions, enabling us to investigate the gain in S/N in a similar way to what is described in the previous sections. 
In this case, we can define the S/N as the fraction between fitted amplitude ($A_\mathrm{IA}b_\mathrm{g}$) and associated error ($\sigma(A_\mathrm{IA}b_\mathrm{g})$) obtained by minimising the $\chi^2$-statistic defined in Eq. (\ref{eq:chi2 transformed}):
\begin{equation}
    \label{eq:SNR fitted scaling}
    \mathrm{S/N}(A_\mathrm{IA}b_\mathrm{g})\coloneqq\frac{A_\mathrm{IA}b_\mathrm{g}}{\sigma(A_\mathrm{IA}b_\mathrm{g})} \,.
\end{equation}
The increase in S/N by adding additional projections is shown in Fig. \ref{Fig:SNR from fit}. 
We show improvement in S/N($A_\mathrm{AI}b_\mathrm{g}$) as a function of added projections for particle number cut (mass cut) in red (blue), both for simple and reduced inertia tensor shapes (points and crosses, respectively). 
Unfilled markers indicate the improvement when going from one to two projections, and filled markers when going from two to three projections. 
Again, the points above the one-to-one line indicate a gain in S/N when adding additional projections. 
As before, increasing the number of projections used increases the S/N. 
The improvement is further quantified in Table \ref{table:SNR gain from fit} where we average over the gain when adding an additional projection. 
Note that with the definition from Eq. (\ref{eq:SNR fitted scaling}), it is no longer necessarily true that the increase in S/N is bounded by the factors $\sqrt{2}$ or $\sqrt{3}$ when adding one or two projections, respectively. We discuss this further below.

\begin{table}

\begin{threeparttable}
\footnotesize
\caption{S/N gain from fit.}
\begin{tabular}{llllll}
\multicolumn{6}{l}{}
\\ \hline\hline
 Stat.   & Shape type & Cut   & 1 vs 3  & 1 vs 2 & 2 vs 3  \\  \hline 
\multirow{4}{*}{$w_\mathrm{g+}$}    & \multirow{2}{*}{simple}    & $n_\star>300$                                  & 1.39   & 1.22             & 1.14           \\  
                                                                        &                                                                    & $M_\star>10^{10.5}M_\odot/h $         & 1.23  & 1.15             & 1.09          \\  
                                                                        & \multirow{2}{*}{reduced}  & $n_\star>300$                                                                             & 1.26     & 1.28             & 1.0          \\  
                                                                        &                                                                         & $M_\star>10^{10.5}M_\odot/h $                                                                                 & 1.46 & 1.4             & 1.05          \\   
\multirow{4}{*}{$\tilde{\xi}_\mathrm{g+}$}    & \multirow{2}{*}{simple}    & $n_\star>300$                                & 1.23     & 1.15             & 1.07           \\  
                                                                        &                                                                    & $M_\star>10^{10.5}M_\odot/h $           & 1.43  & 1.27             & 1.13        \\  
                                                                        & \multirow{2}{*}{reduced}  & $n_\star>300$                                                                                & 1.44   & 1.29             & 1.12         \\  
                                                                        &                                                                         & $M_\star>10^{10.5}M_\odot/h $                                                                                 & 1.51  & 1.43             & 1.07         \\  \hline 
\end{tabular}
\begin{tablenotes}
    \footnotesize
    \item The mean gain in fit amplitude S/N when comparing the use of one versus three; one versus two or two versus three projections.
The gain is shown for $w_\mathrm{g+}$ and $\tilde{\xi}_\mathrm{g+,2}$, for simple and reduced shapes and for both shape samples (mass cuts).
\end{tablenotes}
\label{table:SNR gain from fit}
\end{threeparttable}
\end{table}

Similarly to Sect. \ref{sect:results_SNR}, the S/N increases in almost all cases when adding more projections, whereas the gain is larger when adding a second projection than a third.
A notable exceptions is $w_\mathrm{g+}$ for reduced shapes using the particle number cut, where the S/N decreases slightly on average when adding a third projection. 
We believe this to be cause by the resolution condition (Eq. \ref{eq:cov res}): The error on the correlation function is larger for the higher $r_\mathrm{p}$ we consider in this chapter.
Therefore, adding highly correlated points means proportionally also adding more noise that can lead to removing more values after the SVD; see also Appendix \ref{app:TNG100}.
This problem is unique to noisy covariance matrix estimators that might not have full rank and require the aforementioned treatment.

When considering the full data vector in Sect. \ref{sect:results_SNR}, $w_\mathrm{g+}$ benefits more on average than $\tilde{\xi}_\mathrm{g+}$ when adding additional projections. 
Here, the reverse seems to be the case.
A possible reason for the difference is the additional coupling of the goodness of fit into the S/N consideration. 
Mathematically, the only difference between the $\chi^2$ and the actual S/N is the input model (see Eq. (\ref{eq: SNR_definition}) and (\ref{eq:chi2 transformed}). 
Since the covariance of the fit parameter is closely related to the Hessian of $\chi^2$,\citep{Cramer_1946, Rao_1945}, we expect that improvements in the goodness-of-fit will propagate into the actual S/N($A_\mathrm{AI}b_\mathrm{g}$).
This might further explain why the gain slightly exceeds the theoretical threshold of $\sqrt{2}$ for the reduced-shape $\tilde{\xi}_\mathrm{g+}$ measurements in the high mass cut case, in particular with a noisy jackknife covariance matrix that is limited by the box size. 
The question arises whether this behaviour remains with a more robust covariance that is obtained by analytical means. However, this question is outside the scope of this work and can be tackled in future projects.

Even though the fitted NLA amplitudes are affected by a limited scale range and box size, the S/N shows improvement when combining information of multiple projections in the simulation box. 
The improvement decreasing when going from two to three projections versus from one to two is further in agreement with the results presented in Sect. \ref{sect:results_SNR}. 
This again solidifies the advantages of this method for obtaining the maximum amount of available information for computationally expensive and extensive hydrodynamical cosmological simulations.

\section{Discussion}
\label{sec:discussion}

In this work, we have studied the intrinsic alignments of galaxies in hydrodynamic simulation using the projected shape correlation function, $w_\mathrm{g+}$, and the quadrupole, $\tilde{\xi}_\mathrm{g+,2}$.
Combining the information of the measurements of the intrinsic alignment signals when projected over the three different axis in a simulation box ($x,y,z$), has led to an increased S/N in most studied cases.
Combining two or three projections leads to a gain in S/N for both $w_\mathrm{g+}$ and $\tilde{\xi}_\mathrm{g+,2}$, for shapes measured with the simple or reduced inertia tensor and for two distinct shape samples defined by $n_\star>300$ and $\mathrm{log}(M_\star \ h/\mathrm{M_\odot})>10.5$ (see Fig. \ref{Fig:SNR} and Table \ref{table:SNR gain}).
This gain in S/N when adding a second projection is robust under a change in $\Pi_{\mathrm{max}}$, number of jackknife regions used in the covariance estimate and simulation box size (see appendices \ref{app:pimax}, \ref{app:jk} and \ref{app:TNG100}).
There is additional gain in S/N when adding a third projection in the fiducial case, but this is sensitive to the aforementioned changes explored in Appendices \ref{app:pimax}, \ref{app:jk}, \ref{app:TNG100}, and \ref{app:av cov}, where the gain is more marginal.

Compared to the quadrupole measured in \citet{singh_2024} for TNG100, the $\tilde{\xi}_\mathrm{g+,2}$ signals we measured in TNG300 and TNG100 are very comparable in amplitude and shape.
An in-depth comparison is difficult as the scales probed by \citet{singh_2024} only partially overlap with those we chose to study.
Regarding the $w_\mathrm{g+}$ signal, we find a good qualitative agreement with previous work.
Very similar signal shapes and amplitudes have been measured by e.g. \citet{Samuroff_2021} in TNG300, Illustris and MassiveBlack-II; \citet{Chisari_2015} in HorizonAGN and \citet{Delgado_2023} in MillenniumTNG740, where discrepancies between them mainly arise from differences in galaxy sample selections and the box sizes of the simulations used.
In Fig. \ref{Fig:IA TNG300} we see the effect that sample selection can have on the amplitude of the signal and a similar comparison for TNG100 shows that the same sample definitions will lead to lower signal amplitudes in TNG100 when compared to TNG300.

In agreement with \citep{singh_2024}, comparing the S/Ns of $w_\mathrm{g+}$ to $\tilde{\xi}_\mathrm{g+,2}$, the latter consistently yields a higher S/N for all considered samples, as shown in Fig. \ref{Fig:SNR}.
However, this is strongly dependent on the chosen value of $\Pi_{\mathrm{max}}$, as discussed in Appendix \ref{app:pimax}.
When comparing this to the gain in S/N created by adding the information of multiple projections we see that only when comparing the S/N using one projection for the quadrupole to the S/N using three projections for $w_\mathrm{g+}$, do we find that the S/N(3) for $w_\mathrm{g+}$ is larger than S/N(1) for $\tilde{\xi}_\mathrm{g+,2}$ for a samples using reduced shapes, whereas adding a second projection to $w_\mathrm{g+}$ (S/N(2)) already produces are higher S/N than $\tilde{\xi}_\mathrm{g+,2}$ with one projection, represented by S/N(1), for simple shapes.  
This makes the multipole-based estimator a very promising avenue to extract information about IA models from measurements in a more optimal way, especially when multiple projections are combined.
Future works could explore the effectiveness of combining multiple projections for other estimators, e.g. the one proposed by \citet{lamman2025optimalintrinsicalignmentestimators}, which will likely show similar gains in S/N.

In this work, we model the correlation functions with the non-linear alignment model \citep{Catelan_2001}, that has been shown to be a good model fit for observational galaxy alignment measurements \citep{Singh_Mandelbaum_More_2015, KiDS+DES}. 
Even though the $\chi^2_\mathrm{red}$ indicate good fits for $\tilde{\xi}_\mathrm{g+,2}$ and $w_\mathrm{g+}$, we detect a consistent mis-match of $\sim25\%$ between alignment amplitude obtained from the two different estimators. 
In Appendix \ref{app:fitted amplitude difference} we show that this can be attributed to a different sensitivity of the estimators to nonlinear physics. 
Hence, we expect that not only more conservative scale cuts are necessary for $\tilde{\xi}_\mathrm{g+,2}$ for reliable modelling, but choices of $\Pi_\mathrm{max}$ will influence their comparability as well. 
More stringent restrictions on reliable scales reduce the total S/N of the measurement, possibly diminishing one of the main advantages of using multipole expanded correlation functions instead of projected ones. 
We conclude that, even though $\tilde{\xi}_\mathrm{g+,2}$ has a higher S/N in measurement and corresponding alignment amplitude, better modelling is necessary to exploit the higher S/N of the measurement.
The obtained fitted amplitudes are comparable in magnitude to \citet{singh_2024} who carried out a similar analysis on TNG100 data. 
However, the exact amplitude is highly dependent on galaxy definition and inertia tensor choice, as we demonstrated throughout this work.

Nevertheless, the obtained alignment amplitudes $A_{\mathrm{IA}}b_\mathrm{g}$ are consistent between the different projection directions, both for individual and combined projections. 
By normalising the fitted amplitude with the modelling uncertainty, we obtain S/N($A_\mathrm{AI}b_\mathrm{g}$). 
Similarly to the measurement S/N, we obtain a clear improvement when adding combining projections that subsequently decreases with number of projections. 
It must be noted that the improvement exceeds the theoretical limit $\sqrt{N}$ in the case of reduced shapes with mass cut for $\tilde{\xi}_\mathrm{g+,2}$.
The average improvement when going from one to two projections is slightly higher than $\sqrt{2}$ for this specific measurement.
Furthermore, the S/N gain is smaller when going from one to three projections than from one to two projections for $w_\mathrm{g+}$ in the reduced inertia tensor sample obtained from the particle number cut.
We believe this to be caused by a noisy covariance matrix that propagates into the modelling uncertainty with which we define the S/N($A_\mathrm{AI}b_\mathrm{g}$), in particular since the gain exceeds the threshold for only one of the eight considered cases.
This can be further investigated in the future by using more stable covariance matrix estimators.

While the jackknife resampling technique is widely used for estimating covariance matrices, particularly in IA studies, it has known limitations. 
In the case of this work, it appears that our combined covariance matrix is only stable enough to provide robust results when using the resolution condition, described in Sect. \ref{sect:SNR method}.
In the case of 125 jackknife regions, we filter out 0-6 (mean: 2) elements in the S/N calculation for one projection, 5-13 (7) for two and 9-17 (11) for three, where more elements are filtered out for the quadrupole than $w_\mathrm{g+}$ and simple than reduced shapes; and none of the elements that are used in fitting are filtered out.
For 64 jackknife regions, in each case, more elements are filtered out than for 125 jackknife regions.
Furthermore, we have seen in Appendix \ref{app:jk} that the results are sensitive to the number of jackknife regions used.
To further investigate this, it would be valuable to compare our jackknife covariance matrix to an analytically calculated covariance matrix. 
This is a complicated task but presents a possibility for future research. 
Taking the example of \citet{Samuroff_2021}, they compare their jackknife covariance matrix to an analytically calculated matrix and find that on scales $r_\mathrm{p} > 6$\,Mpc$/h$, the jackknife tends to underestimate the variance of $w_\mathrm{g+}$ by up to $\sim$ 25-50 percent for the TNG300 simulation. 

Another viable avenue of further research would include being able to predict covariance of the analytical relations between the projected shapes of our measured ellipsoids.
While this is outside of the scope of this work, it would provide insights into why we see a gain in information when adding multiple projections.

As many hydrodynamical simulations currently in use will have comparable or larger box sizes than TNG300, we can assume that our results can be used to measure the IA signals with a consistently larger S/N.
Using relatively little computing power compared to the amount needed to run larger simulations, we can use the existing hydrodynamical simulations to better constrain modelling priors currently derived from these same simulations.

\section{Conclusions}
In this work, we have combined the information of shapes of galaxies projected over multiple axes, while taking their covariance into account, in hydrodynamical simulation TNG300.
We show that a consistent and robust gain in S/N can be obtained for both $w_\mathrm{g+}$ and $\tilde{\xi}_\mathrm{g+,2}$ by combining multiple projections.
Below, the main conclusions are summarised.
   \begin{enumerate}
      \item Combining the information of more than one projection, the S/Ns of both $w_\mathrm{g+}$ and $\tilde{\xi}_\mathrm{g+,2}$ are consistently larger than for a single projection (Table \ref{table:SNR gain} and Fig. \ref{Fig:SNR}).
      \item Adding a second projection leads to a larger gain in S/N than adding a third projection for both $w_\mathrm{g+}$ and $\tilde{\xi}_\mathrm{g+,2}$ (Table \ref{table:SNR gain} and Fig. \ref{Fig:SNR}).
      \item The gain in S/N when adding two or three projections is larger for shapes measured with the reduced inertia tensor than with the simple inertia tensor (Table \ref{table:SNR gain} and Fig. \ref{Fig:SNR}).
      \item Table \ref{table:SNR gain} shows that the gain in S/N for a high mass shape sample ($\mathrm{log}(M_\star \ h/\mathrm{M_\odot})>10.5$) is similar to that for all galaxies with resolved shapes ($n_\star>300$), where the high mass sample signal has a much higher amplitude (Fig. \ref{Fig:IA TNG300}).
      \item The S/N of the fitted amplitude obtained from the non-linear alignment model increases when adding projections in almost all cases but shows inconclusive systematic behaviour between estimators and sample selection. This is likely caused by the increased uncertainty on larger scales commonly used for modelling (Table \ref{table:SNR gain from fit} and Fig. \ref{Fig:SNR from fit}).
   \end{enumerate}

\begin{acknowledgements}
We thank Casper Vedder for plenty of useful discussions. This publication is part of the project ``A rising tide: Galaxy intrinsic alignments as a new probe of cosmology and galaxy evolution'' (with project number VI.Vidi.203.011) of the Talent programme Vidi which is (partly) financed by the Dutch Research Council (NWO). DN and HH acknowledge support from the European Research Council (ERC) under the European Union’s Horizon 2020 research and innovation program with Grant agreement No. 101053992.\end{acknowledgements}

\bibliographystyle{aa}
\bibliography{refs}

\begin{appendix}

\section{$\Pi_{\mathrm{max}}$}
\label{app:pimax}

\begin{figure} 
    \centering
    \includegraphics[width=0.5\textwidth]{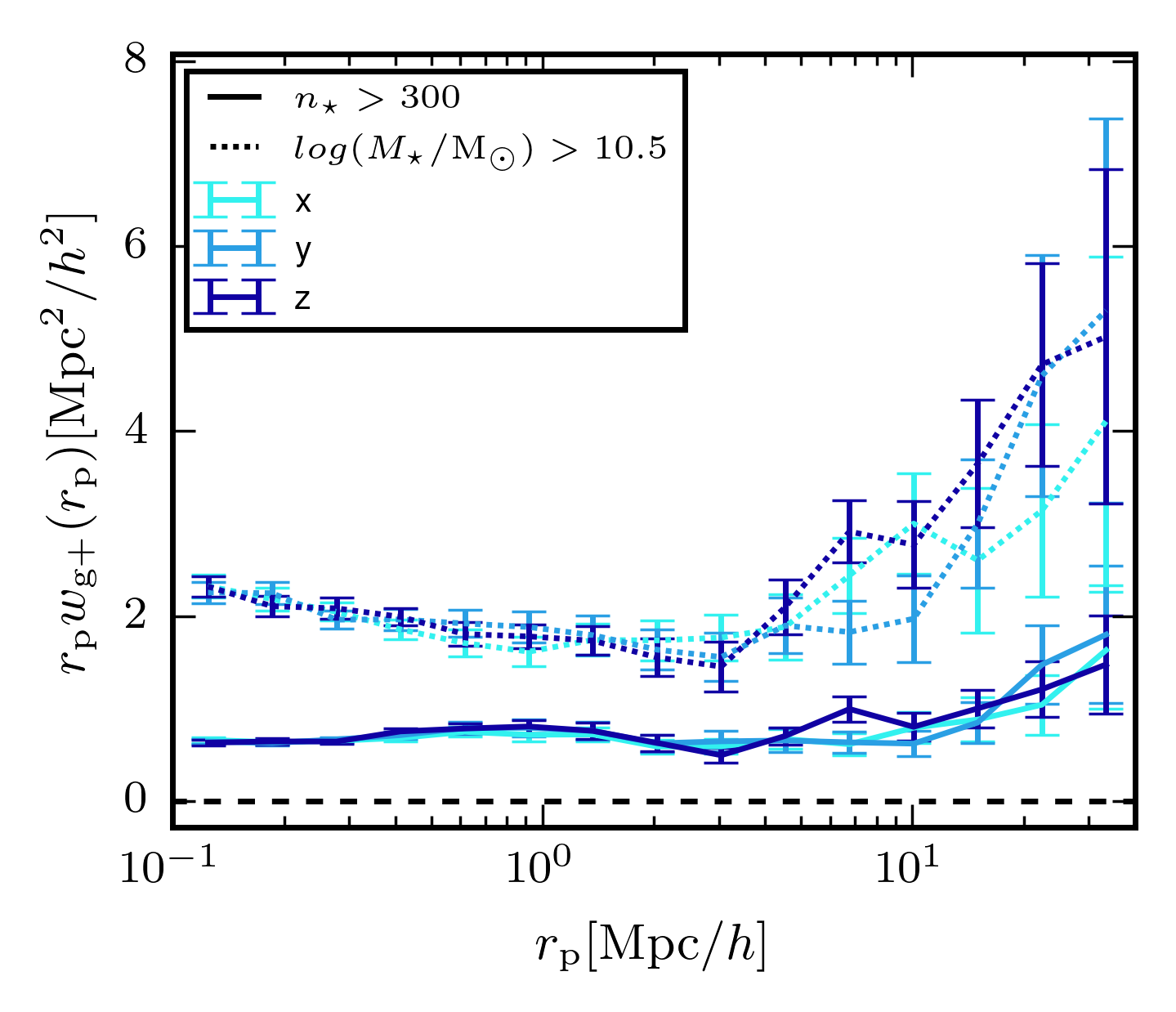}
    \caption{Correlation function, $r_p w_\mathrm{g+}$, for $\Pi_{\mathrm{max}}=102.5\mathrm{Mpc}/h$, in TNG300 for two shape samples: $n_\star>300$ (continuous lines) and $\mathrm{log}(M_\star \ h/\mathrm{M_\odot})>10.5$ (dashed lines).
    The shapes are projected over the $x$ (light blue), $y$ (medium blue) and $z$ (dark blue) axes and measured using the simple inertia tensor.
    }
    \label{Fig:IA TNG300 pimax}
\end{figure}

As mentioned in Sect. \ref{sect: wg+ method}, the choice in the integration limit $\Pi_{\mathrm{max}}$ can have a significant impact on the measurement of $w_\mathrm{g+}$.
In order to match the measurements of the quadrupole more closely, which is dependent on the 3D separation between a galaxy pairs, the results in the main body of the text are measured with $\Pi_{\mathrm{max}}=20\mathrm{Mpc}/h$, matching the maximum $r$ and $r_p$ values.
In order to give so insight into the influence of the choice in $\Pi_{\mathrm{max}}$ on the results, Fig. \ref{Fig:IA TNG300 pimax} shows $r_p w_\mathrm{g+}$ when measured with $\Pi_{\mathrm{max}}=102.5\mathrm{Mpc}/h$, for both shape samples $n_\star>300$ (continuous lines) and $\mathrm{log}(M_\star \ h/\mathrm{M_\odot})>10.5$ (dashed lines) in all three projections $x$ (light blue), $y$ (medium blue) and $z$ (dark blue), similarly to Fig. \ref{Fig:IA TNG300} in the main text.

When comparing Fig. \ref{Fig:IA TNG300 pimax} to the top panel of Fig. \ref{Fig:IA TNG300}, we can see that the shape of the correlation functions has changed, mainly for large scales ($r_p\gtrsim3\mathrm{Mpc}/h$).
The signal shape matches that of the quadrupole (Fig. \ref{Fig:IA TNG300}, bottom panel) less closely.
Moreover, the signal at these larger scales is much noisier for a larger $\Pi_{\mathrm{max}}$, which is apparent in both the larger error bars in Fig. \ref{Fig:IA TNG300 pimax} and the divergence of the different projections.
\FloatBarrier

\begin{table}
\begin{threeparttable}
\footnotesize
 \caption{S/N gain $\Pi_{\mathrm{max}}=102.5\mathrm{Mpc}/h$.}
\begin{tabular}{llllll}
\multicolumn{6}{l}{}
\\ \hline\hline
 Stat.    & Shape type & Cut   & 1 vs 3  & 1 vs 2 & 2 vs 3  \\  \hline 
\multirow{4}{*}{$w_\mathrm{g+}$}    & \multirow{2}{*}{simple}    & $n_\star>300$                                  & 1.26   & 1.2             & 1.05           \\  
                                                                        &                                                                    & $M_\star>10^{10.5}M_\odot/h $         & 1.21  & 1.18             & 1.03          \\  
                                                                        & \multirow{2}{*}{reduced}  & $n_\star>300$                                                                             & 1.41     & 1.3             & 1.09          \\  
                                                                        &                                                                         & $M_\star>10^{10.5}M_\odot/h $                                                                                 & 1.29 & 1.2             & 1.07          \\  \hline 
\end{tabular}
\begin{tablenotes}
\footnotesize
\item 
The mean gain in S/N when comparing the use of one versus three; one versus two or two versus three projections.
The gain is shown for $w_\mathrm{g+}$, for simple and reduced shapes and for both shape samples, using $\Pi_{\mathrm{max}}=102.5\mathrm{Mpc}/h$ (compare to Table \ref{table:SNR gain}).
\end{tablenotes}
\label{table:SNR pimax}
\end{threeparttable}
\end{table}

\begin{figure}
    \centering
    \includegraphics[width=0.5\textwidth]{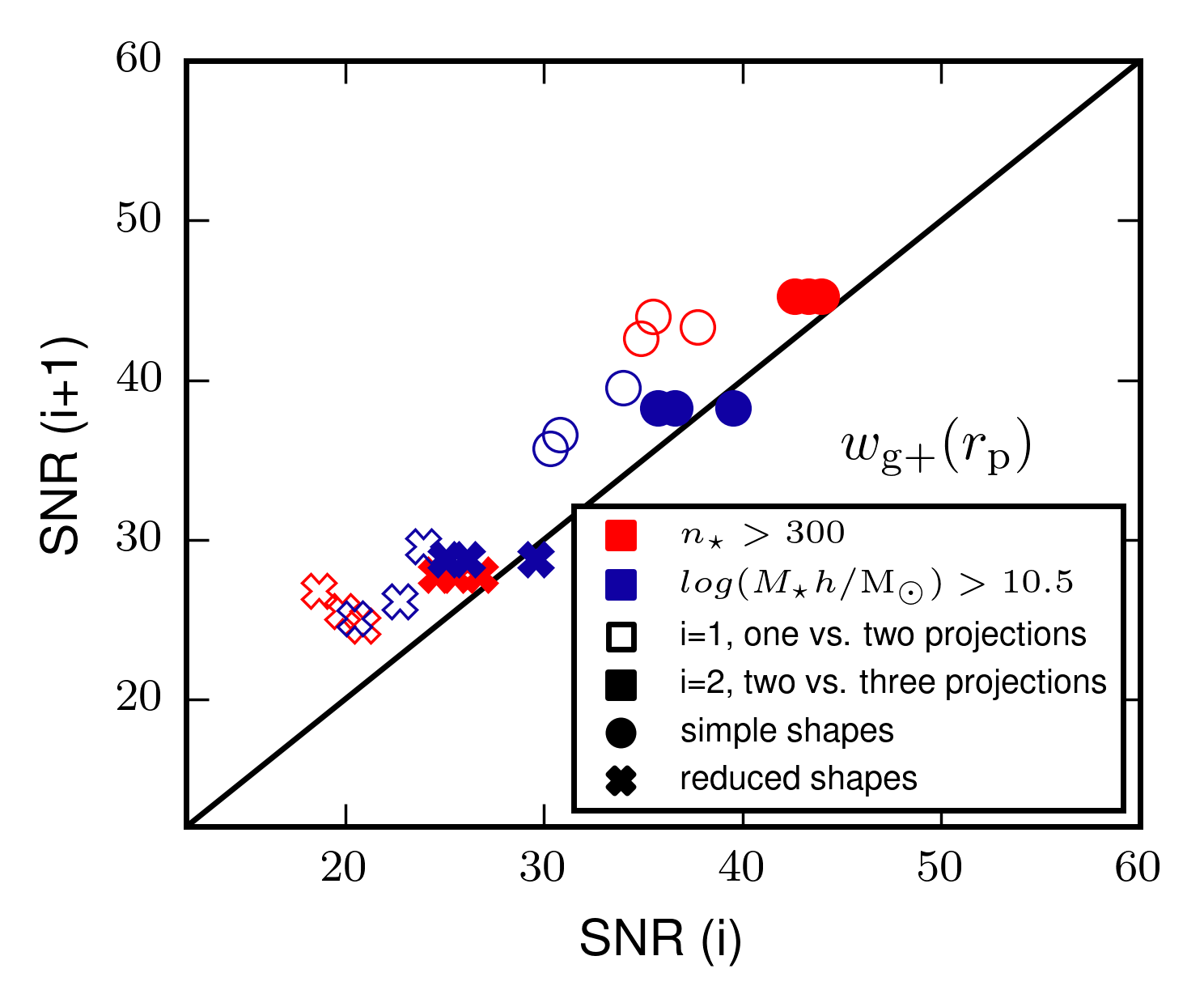}
    \caption{
    S/Ns for the $w_\mathrm{g+}$ of combining one vs two (unfilled) and two vs three (filled) projections.
    The colours of the markers correspond to the different shape samples: $n_\star>300$ is red and $\mathrm{log}(M_\star \ h/\mathrm{M_\odot})>10.5$ is blue.
    Shapes measured using the simple, reduced inertia tensor are depicted by circles, crosses, respectively.
    The black line denotes the boundary for which both S/Ns are equal.
    In this figure, $\Pi_{\mathrm{max}}=102.5\mathrm{Mpc}/h$, instead of $20\mathrm{Mpc}/h$ as it is in Fig. \ref{Fig:SNR}.
    }
    \label{Fig:SNR pimax}
\end{figure}

Figure \ref{Fig:SNR pimax} shows the S/Ns of one versus two (unfilled markers) projections and two versus three (filled) for simple (circles) and reduced (crosses) shapes and for both mass cuts $n_\star>300$ (red) and $\mathrm{log}(M_\star h/\mathrm{M_\odot})>10.5$ (blue).
The figure is equivalent to the top panel of Fig. \ref{Fig:SNR} in the main text, with the only difference being that here $w_\mathrm{g+}$ was calculated using a $\Pi_{\mathrm{max}}$ of $102.5\mathrm{Mpc}/h$.
Comparing Fig. \ref{Fig:SNR pimax} to the top panel of Fig. \ref{Fig:SNR}, we can see that a larger $\Pi_{\mathrm{max}}$ results in much lower S/Ns for $w_\mathrm{g+}$ because of the noisier data vector.
Looking at both Fig. \ref{Fig:SNR pimax} and Table \ref{table:SNR pimax}, we see that the rest of the conclusions from the main text remain valid: adding a second projection has a larger gain than the third, although each projection added increases the S/N; the gain in S/N is higher for reduced shapes than simple ones and the gains are comparable between the mass cuts.
However, comparing Table \ref{table:SNR pimax} to Table \ref{table:SNR gain}, we see that the gains in S/N when adding projections are higher for a larger value of $\Pi_{\mathrm{max}}$, even though the S/Ns themselves remain lower.

\section{Jackknife regions}
\label{app:jk}

\begin{table}
\begin{threeparttable}
\footnotesize
 \caption{S/N gain for 64 jackknife regions.}
\begin{tabular}{llllll}
\multicolumn{6}{l}{}
\\ \hline\hline
 Stat.    & Shape type & Cut   & 1 vs 3  & 1 vs 2 & 2 vs 3  \\  \hline 
\multirow{4}{*}{$w_\mathrm{g+}$}    & \multirow{2}{*}{simple}    & $n_\star>300$                                  & 1.09   & 1.1             & 0.99           \\  
                                                                        &                                                                    & $M_\star>10^{10.5}M_\odot/h $         & 1.08  & 1.21             & 0.89          \\  
                                                                        & \multirow{2}{*}{reduced}  & $n_\star>300$                                                                             & 1.12     & 1.18             & 0.96          \\  
                                                                        &                                                                         & $M_\star>10^{10.5}M_\odot/h $                                                                                 & 1.19 & 1.27             & 0.95          \\   
\multirow{4}{*}{$\tilde{\xi}_\mathrm{g+}$}    & \multirow{2}{*}{simple}    & $n_\star>300$                                & 1.05     & 1.07             & 0.98           \\  
                                                                        &                                                                    & $M_\star>10^{10.5}M_\odot/h $           & 1.05  & 1.13             & 0.93        \\  
                                                                        & \multirow{2}{*}{reduced}  & $n_\star>300$                                                                                & 1.25   & 1.18             & 1.07         \\  
                                                                        &                                                                         & $M_\star>10^{10.5}M_\odot/h $                                                                                 & 1.06  & 1.12             & 0.95         \\  \hline 
\end{tabular}
\begin{tablenotes}
\footnotesize
\item 
The mean gain in S/N when comparing the use of one versus three; one versus two or two versus three projections.
The gain is shown for $w_\mathrm{g+}$ and $\tilde{\xi}_\mathrm{g+,2}$, for simple and reduced shapes and for both shape samples, using 64 jackknife regions (compare to Table \ref{table:SNR gain}).
\end{tablenotes}
\label{table:SNR jk64}
\end{threeparttable}
\end{table}

\begin{figure*}
    \begin{center}
    \centering
    \includegraphics[width=\textwidth]{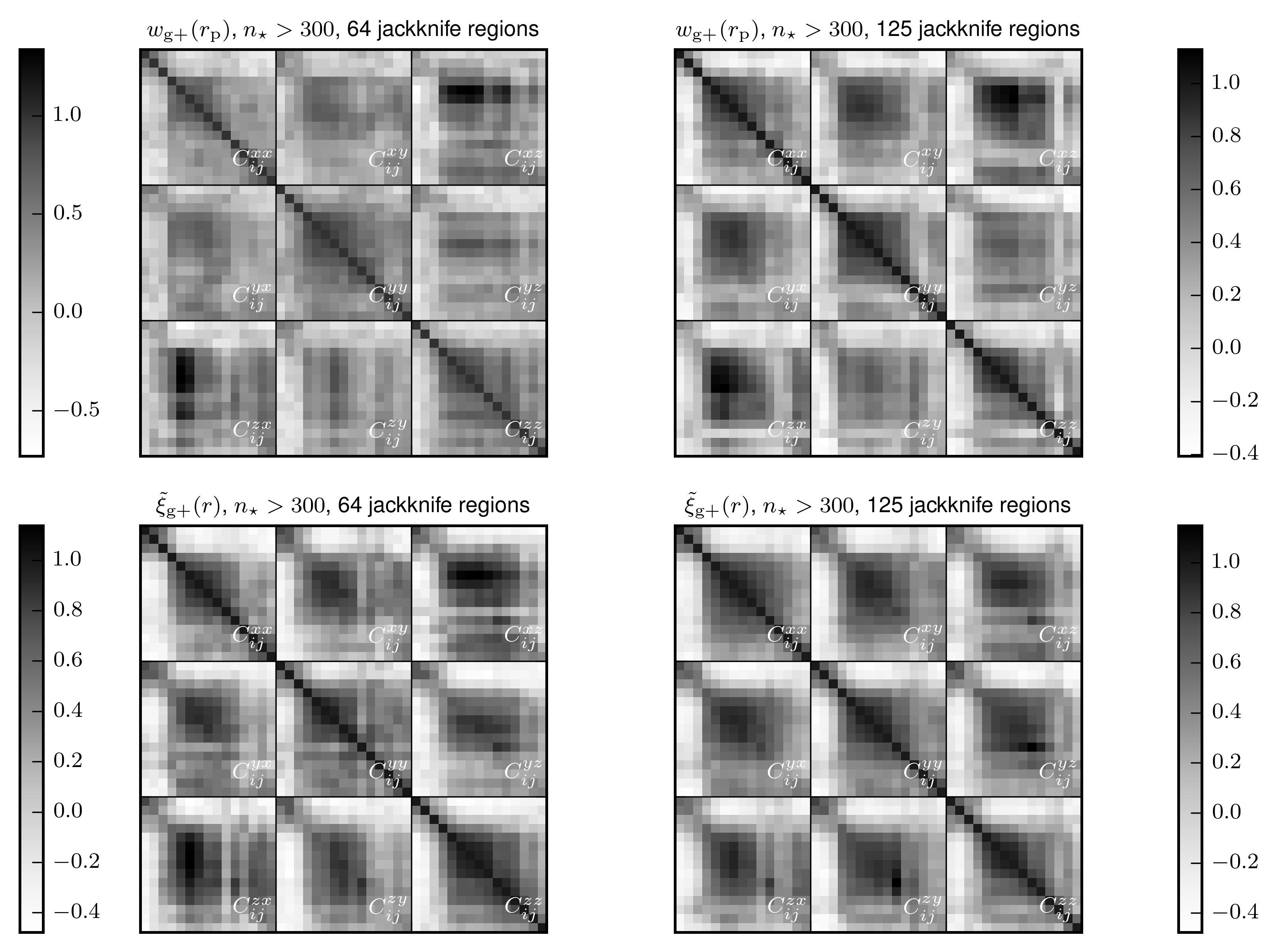}
    \caption{Normalised combined covariance matrices for $w_\mathrm{g+}$ (top) and $\tilde{\xi}_\mathrm{g+,2}$ (bottom) for the $n_\star>300$ shape sample, using 64 jackknife samples (left) or 125 jackknife regions (right).
    Darker colours indicate higher values.
    Each block of the combined covariance matrix is labelled according to the definitions in Sect. \ref{sect:cov}. The right side of the figure is equal to Fig. \ref{Fig:covariancematrix}.
    }
    \label{Fig:covariancematrix 64jk}
    \end{center}
\end{figure*}

The chosen number of jackknife regions can potentially have a large impact on the S/Ns obtained, as the over- or underestimation of the covariance can lead to very different S/Ns.
Furthermore, if the choice in the number of jackknife regions leads to instability in the covariance, the resulting S/Ns can become unreliable and even nonsensical.
Especially for the combination of three projections, the covariance matrix is sensitive to these instabilities, due to the high number of strongly correlated elements.
Therefore, we have studied two options for the number of jackknife regions: 125 and 64, settling on 125 for TNG300, which is used in the main body of the text as it seems to be the most stable.

The superior stability of 125 jackknife regions versus 64 can be substantiated in two ways.
First, when applying the covariance resolution condition, as described in Sect. \ref{sect:SNR method}, we see that in the case of 125 jackknife regions, systematically, more eigenvalues of the SVD pass the resolution criterium and therefore contribute to the S/N than for 64 jackknife regions.
Second, we expect the addition of each term to result in a monotonic increase in the S/N.
While this is not true in every case, we see that it is true for many more of the cases using 125 jackknife regions than 64 regions.
Combined with the invertability of the matrices and the reasonable values obtained for the S/Ns, the choice of 125 jackknife regions seems to make the most sense.

In this appendix the normalised combined covariance matrices for $w_\mathrm{g+}$ and $\tilde{\xi}_\mathrm{g+,2}$ are shown for 64 jackknife regions ($4^3$, left side) and 125 jackknife regions ($5^3$, right side) in Fig. \ref{Fig:covariancematrix 64jk}, where the right side is is analogous to Fig. \ref{Fig:covariancematrix} and has been added for comparison.

As mentioned in Sect. \ref{sect:cov results}, the values of the normalised combined covariance are lower when using 64 jackknife regions than 125, especially for $w_\mathrm{g+}$.
This can be seen in Fig. \ref{Fig:covariancematrix 64jk}, when comparing the left and right side of the figure: the right side is generally darker, indicating higher values.
When taking the median of the ratios between 64 and 125 regions for all elements, we find that this is $\sim0.95$ for $w_\mathrm{g+}$ and $\sim1.0$ for $\tilde{\xi}_\mathrm{g+,2}$.
Barring that, the same basic structures arise in both versions of the covariance, consolidating the trust in the measurement accuracy in general.

\begin{figure}
    \centering
    \includegraphics[width=0.5\textwidth]{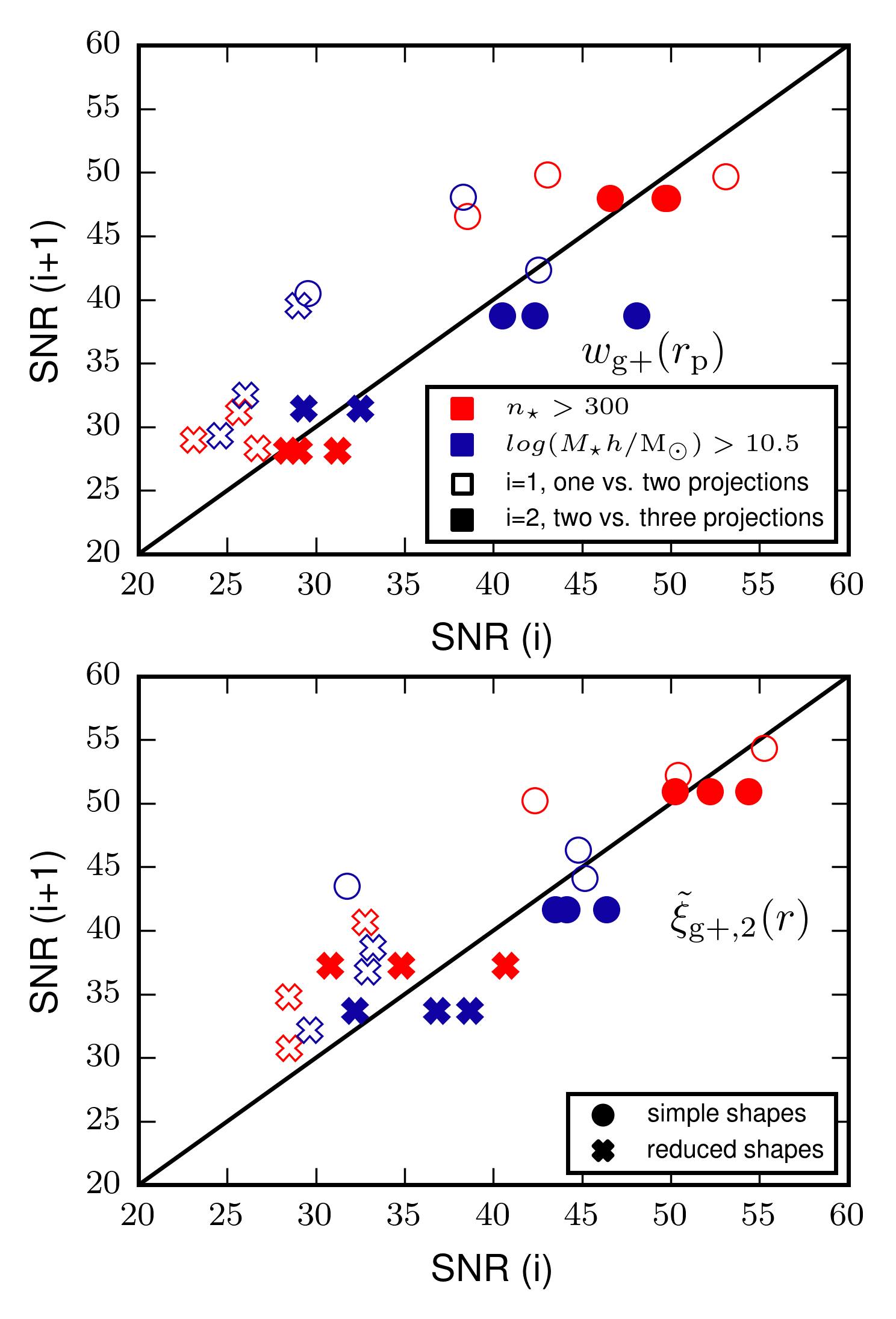}
    \caption{
    S/Ns for the $w_\mathrm{g+}$ (top) and $\tilde{\xi}_\mathrm{g+,2}$ (bottom) of combining one versus two (unfilled) and two versus three (filled) projections in TNG300 for 64 jack-knife regions.
    The colours of the markers correspond to the different shape samples: $n_\star>300$ is red and $log(M_\star h/\mathrm{M_\odot})>10.5$ is blue.
    Shapes measured using the simple, reduced inertia tensor are depicted by circles, crosses, respectively.
    The black line denotes the boundary for which both S/Ns are equal.
    }
    \label{Fig:SNR jk}
\end{figure}

The impact of the number of jackknife regions on the S/N gains can be seen in Fig. \ref{Fig:SNR jk} and Table \ref{table:SNR jk64}.
While decreasing the number of jackknife regions alters the S/N compared to Fig. \ref{Fig:SNR} and Table \ref{table:SNR gain}, the main conclusions presented in Sect. \ref{sect:results_SNR} remain valid.
Adding the second projections increases the S/N in nearly all cases, and in some, adding a third projection also adds additional S/N.
The S/N gain also seems to be higher for reduced shapes than for the simple ones and is once again comparable for both shape samples (mass cuts).
Furthermore, the difference gain in $w_\mathrm{g+}$ and $\tilde{\xi}_\mathrm{g+,2}$ seen in Sect. \ref{sect:results_SNR} is also still there.

The only qualitative difference between using 64 or 125 jackknife regions for the S/N gain, is that when using 64 regions, there are many cases (mostly for simple shapes) where adding a third projection is decreasing the S/N instead of increasing it, leading to sub-one ratios in Table \ref{table:SNR jk64}.

In conclusion, the main trends and conclusions seem trustworthy, although there is some sensitivity of the results to the number of jackknife regions.
This mostly shows the limitations of the jackknife method for covariance estimation, even when correcting for biases.
Ideally, we would use multiple independent realisations of the simulation to estimate the covariance, but with the cost of running hydrodynamical simulations, this is not feasible.

\section{Box size and resolution}
\label{app:TNG100}

This appendix explores the influence of the simulation box size and resolution on the gain in S/N when adding the information of two or three projections to the intrinsic alignment correlation functions.
These effects are explored together by doing the same measurements as in the rest of the paper for TNG100-1 instead of TNG300-1, which has a smaller box size ($L_{box}=75 c\mathrm{Mpc}/h$ and a higher resolution ($1820^3$ dark matter particles of mass $m_{\mathrm{DM}}=7.5 \times 10^6 \mathrm{M_\odot}$).

\begin{figure}
    \centering
    \includegraphics[width=0.5\textwidth]{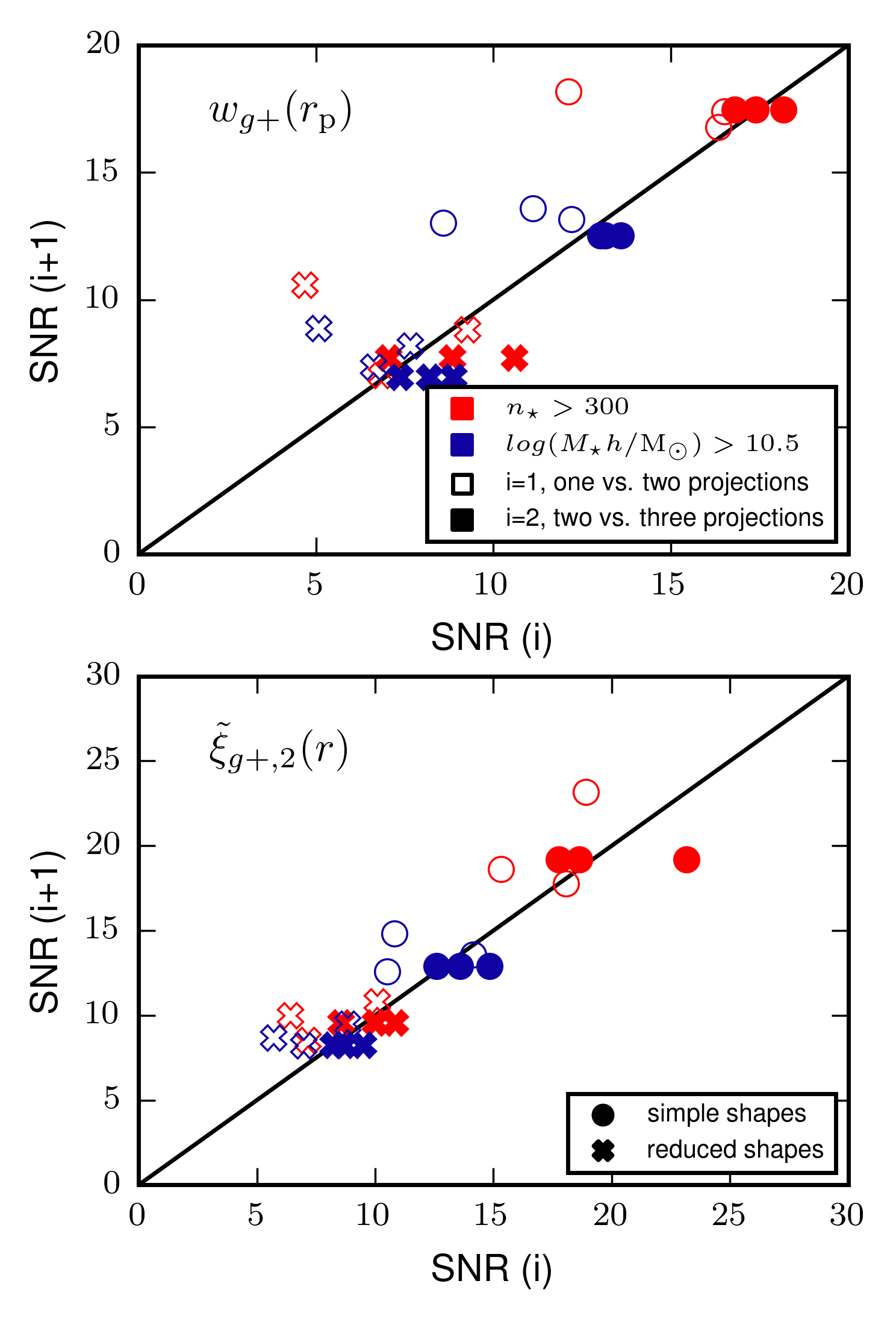}
    \caption{
    S/Ns for the $w_{g+}$ (top) and $\tilde{\xi}_{g+,2}$ (bottom) of combining one versus two (unfilled) and two versus three (filled) projections in TNG100.
    The colours of the markers correspond to the different shape samples: $n_\star>300$ is red and $log(M_\star h/\mathrm{M_\odot})>10.5$ is blue.
    Shapes measured using the simple, reduced inertia tensor are depicted by circles, crosses, respectively.
    The black line denotes the boundary for which both S/Ns are equal.
    }
    \label{Fig:SNR TNG100}
\end{figure}

Figure \ref{Fig:SNR TNG100} and Table \ref{table:SNR TNG100} are the counterparts of Fig. \ref{Fig:SNR} and Table \ref{table:SNR gain}, showing the gain in S/N when adding projections in TNG100 for the two shape samples and shape types in both $w_{g+}$ (top panel) and $\tilde{\xi}_{g+,2}$ (bottom).
As the box size of TNG100 is smaller, we opted to use 64 jackknife regions to measure the covariance and used $15 \ r_{(\mathrm{p})}$ bins between $0.1 \,\mathrm{Mpc}/h$ and $20 \,\mathrm{Mpc}/h$.
Figure \ref{Fig:SNR TNG100}, shows that while the markers lie much closer to the line than in Fig. \ref{Fig:SNR} (for TNG300), the unfilled markers generally still lie in the top left half of the figure, indicating a gain in S/N when adding a second projection (going from one to two projections: unfilled markers; from two to three: filled).
The figure axes limits announce that we are in a difference S/N regime: the S/Ns in TNG100 are much lower than in TNG300.
This is to be expected, as the box size is much smaller and the measurements are therefore noisier.
Adding the information presented in Table \ref{table:SNR TNG100}, we see that our main conclusions, exhibited in Sect. \ref{sect:results_SNR}, hold.
As in TNG300, in TNG100 the gains in S/N are larger when adding the second to the first projection than adding the third to the second (if there is any gain in when adding the third).
Also, the gains in S/N when adding projections for the reduced shapes (crosses) are higher than those for the simple shapes (circles) more often than not.
Furthermore, the gains in S/N when adding projections (and the S/N values themselves) for the shape samples defined by $n_\star>300$ (red) and $log(M_\star h/\mathrm{M_\odot})>10.5$ (blue) are comparable.

\FloatBarrier

\begin{table}
\begin{threeparttable}
\footnotesize
\caption{S/N gain TNG100.}
\begin{tabular}{llllll}
\multicolumn{6}{l}{}
\\ \hline\hline
 Stat.    & Shape type & Cut   & 1 vs 3  & 1 vs 2 & 2 vs 3  \\  \hline 
\multirow{4}{*}{$w_{\mathrm{g}+}$}    & \multirow{2}{*}{simple}    & $n_\star>300$                                  & 1.19   & 1.19             & 1.0           \\  
                                                                        &                                                                    & $M_\star>10^{10.5}M_\odot/h $         & 1.2  & 1.27             & 0.94          \\  
                                                                        & \multirow{2}{*}{reduced}  & $n_\star>300$                                                                             & 1.2     & 1.42             & 0.9          \\  
                                                                        &                                                                         & $M_\star>10^{10.5}M_\odot/h $                                                                                 & 1.11 & 1.31             & 0.86          \\   
\multirow{4}{*}{$\tilde{\xi}_{\mathrm{g}+}$}    & \multirow{2}{*}{simple}    & $n_\star>300$                                & 1.11     & 1.14             & 0.98           \\  
                                                                        &                                                                    & $M_\star>10^{10.5}M_\odot/h $           & 1.11  & 1.18             & 0.95        \\  
                                                                        & \multirow{2}{*}{reduced}  & $n_\star>300$                                                                                & 1.26   & 1.28             & 0.99         \\  
                                                                        &                                                                         & $M_\star>10^{10.5}M_\odot/h $                                                                                 & 1.19  & 1.26             & 0.94         \\  \hline 
\end{tabular}
\begin{tablenotes}
\footnotesize
\item The mean gain in S/N when comparing the use of one versus three; one versus two or two versus three projections.
The gain is shown for $w_{g+}$ and $\tilde{\xi}_{g+,2}$, for simple and reduced shapes and for both shape samples, for TNG100 (compare to Table \ref{table:SNR gain}).
\end{tablenotes}
\label{table:SNR TNG100}
\end{threeparttable}
\end{table}

In conclusion, the main conclusions drawn in Sect. \ref{sect:results_SNR} are not severely impacted by simulation box size and resolution.
While individual cases will differ, the trends are robust.
Combining the results in this appendix, to those in Appendices \ref{app:pimax} and \ref{app:jk} we can state that when the measurement becomes noisier (due to e.g. higher $\Pi_{\mathrm{max}}$, fewer jackknife regions or smaller box size), the impact of adding a second projection is increased, while the impact of adding a third projection is mitigated by the addition of extra noise.

\section{Motivation of the resolution condition in given by Eq. (\ref{eq:cov res})}
\label{app: Derivation of the resolution cut}

We first remind ourselves that the Jackknife covariance estimate is given by Eq. (\ref{Cjk: simple}). Here, we define the difference between Jackknife per-region estimate and mean as $\Delta_i^{[k]}\coloneqq\psi_i^k - \Bar{\psi}_i$, with $\Bar{\psi}_i$ from Eq. (\ref{eq: Jackknife_bar_estimate}). Therefore, the Jackknife covariance estimate becomes

\begin{equation}\label{Cjk: simple_appendix}
    \hat{C}_{ij} = \frac{N_\mathrm{sub}-1}{N_\mathrm{sub}} \sum_{k=1}^{N_\mathrm{sub}} \Delta_i^{[k]}\Delta_j^{[k]}.
\end{equation}

We stress that in this section $\hat{C}_{ij}$ denotes the covariance estimator and not the rescaled covariance matrix described in Sect. \ref{sect:SNR method}. 
The individual $\left\{\{\Delta_i^{[k]}\}_{i=1\ldots p}\right\}_{k=1\ldots N_\mathrm{sub}}$ are matrix elements for the $p\times  N_\mathrm{sub}$ random matrix, where $N_\mathrm{sub}$ is the number of Jackknife regions and $p$ the number of used $r_{\mathrm{(p)}}$ bins, i.e. measurement points. 
Assuming the points to be drawn from a $p$-dimensional Gaussian $\mathcal{N}(0,\mathbf{\Sigma'})$ with scale matrix $\mathbf{\Sigma'}$, Eq. (\ref{Cjk: simple_appendix}) is  of the same functional form as the Wishart-distributed random scatter matrix\,\citep{Wishart_1928,gupta2000matrix}
\begin{equation}
    \mathcal{S}_{ij}=\sum_{k=1}^N \Delta_i^{[k]}\Delta_j^{[k]}.
\end{equation}
Denoting $\mathbf{\Sigma}$ as the true covariance, the variance of $\hat{C}_{ij}$ then directly follows from the Wishart distribution using $\nu=N_\mathrm{sub}-1$ degrees of freedom and the scale matrix $\mathbf{\Sigma'}=\mathbf{\Sigma}/(N_\mathrm{sub}-1)$ \citet{Mohammad2022Jackknife} as\footnote{Note that the expectation value of the Wishart distribution fulfills $\mathbb{E}(\mathcal{S})=\nu\mathbf{\Sigma'}$, hence we choose $\mathbf{\Sigma}'=\mathbf{\Sigma}/\nu$ to obtain the true covariance as expectation.}
\begin{align}
    \mathrm{Var}(\hat{C}_{ij}) & = \frac{N_\mathrm{sub}-1}{N_\mathrm{sub}} \nu \left((\Sigma'_{ij})^2+\Sigma'_{ii}\Sigma'_{ij}\right) \nonumber \\
    & = \frac{1}{N_\mathrm{sub}}\left((\Sigma_{ij})^2+\Sigma_{ii}\Sigma_{ij}\right)\,. 
\end{align}
We note that we would obtain the expression given in \citet{Gaztanaga_2005}, $\Delta C_{ij}\simeq\sqrt{\frac{2}{N_\mathrm{sub}}}$, by taking the limiting case of $i=j$ and $|\Sigma_{ij}|\simeq\mathcal{O}(1)$. 
The latter holds if we perform the rescaling as specified in Eq. (\ref{eq:cov norm}). 
The former provides an upper bound under the reasonable assumption that the off-diagonals of the covariance matrix are smaller than the diagonal entries.

In the case of a positive-semidefinit real matrix (e.g. a covariance), the SVD is given by
\begin{equation}
    \boldsymbol{\hat{C}=U}^\mathrm{T}\boldsymbol{D\,U} \qquad \mathrm{with} \qquad \boldsymbol{U}^\mathrm{T}\boldsymbol{U=I}\,.
\end{equation}
Thus, we can translate the bound on $\Delta C_{ij}$ to a cut for eigenvalues after SVD by splitting the estimated covariance into a signal and noise term 
\begin{align}
    \boldsymbol{\hat{C}}&=\boldsymbol{\Sigma}+\boldsymbol{N}=\boldsymbol{\hat{U}}^\mathrm{T}\boldsymbol{\hat{D}}\,\boldsymbol{\hat{U}} \nonumber \\
    \Longrightarrow \qquad \boldsymbol{\hat{D}} & = \boldsymbol{\hat{U}\,\Sigma\,\hat{U}}^\mathrm{T} +\boldsymbol{\hat{U}\,N\,\hat{U}}^\mathrm{T} \simeq \boldsymbol{D}+\boldsymbol{\hat{U}\,N\,\hat{U}}^\mathrm{T} \,.
\end{align}
For the individual SVD eigenvalues, this which yields {$\hat{\lambda}_i^2 \simeq \lambda_i^2 + \hat{U}_i\boldsymbol{N}\,\hat{U}_i^\mathrm{T}.$}
The expected noise contribution is then\begin{align}
    \mathrm{Var}(\hat{U}_i\boldsymbol{N}\,\hat{U}_i^\mathrm{T})&=\mathrm{Var}\left(\sum_{j,k}\hat{U}_{ij}N_{jk}\hat{U}_{ki}\right)\nonumber\\
    &=\sum_{j,k}\hat{U}_{ij}\mathrm{Var}(N_{jk})\hat{U}_{ki}\nonumber \\
    & = \sum_{j,k}\hat{U}_{ij}(\Delta C_{jk})^2\hat{U}_{ki}\nonumber\\
    &\simeq \frac{2}{N_\mathrm{sub}}\sum_{jk}\hat{U}_{ij}\hat{U}_{ki}=\frac{2}{N_\mathrm{sub}}\,,
\end{align}
where we used the orthogonality of $\boldsymbol{\hat{U}}$. Therefore, $\sigma_i\simeq\sqrt{\frac{2}{N_\mathrm{sub}}}$ and we expect singular values $\hat{\lambda}_i^2<\sqrt{\frac{2}{N_\mathrm{sub}}}$ to be dominated by noise.

\section{Averaged covariance}
\label{app:av cov}

The full covariance matrix containing all three projections (Sect. \ref{sect:cov results}, Fig. \ref{Fig:covariancematrix}) shows unexpected discrepancies between the off-diagonal blocks and the diagonal blocks, which should be equal in a perfect world.
For a large enough simulation box, we expect there to be no preferred line-of-sight direction, which means that $C_{ij}^{xx}=C_{ij}^{yy}=C_{ij}^{zz}$ and $C_{ij}^{xy}=C_{ij}^{xy}=C_{ij}^{yz}$.
The current discrepancies can be attributed to the noisiness of the jackknife covariance estimator and the influence of variations in the large scale structure due to the limited box size.
Averaging the diagonal blocks and the off-diagonal blocks, and using this averaged covariance matrix to calculate the S/Ns, gives a lower bound to the S/N gain.
This procedure produces a lower bound because the differences in the large scale structure between projections are ignored, which should be taken into account, as they are non-trivial in small simulation boxes.
\FloatBarrier
\begin{table}
\begin{threeparttable}
\footnotesize
\caption{S/N gain for averaged covariance.}
\begin{tabular}{llllll}
\multicolumn{6}{l}{}
\\ \hline\hline
Stat. & Shape type & Cut   & 1 vs 3  & 1 vs 2 & 2 vs 3  \\  \hline 
\multirow{4}{*}{$w_\mathrm{g+}$}    & \multirow{2}{*}{simple}    & $n_\star>300$                                  & 1.1   & 1.12             & 0.98           \\  
                                                                        &                                                                    & $M_\star>10^{10.5}M_\odot/h $         & 1.16  & 1.14             & 1.02          \\  
                                                                        & \multirow{2}{*}{reduced}  & $n_\star>300$                                                                             & 1.16     & 1.16             & 1.0          \\  
                                                                        &                                                                         & $M_\star>10^{10.5}M_\odot/h $                                                                                 & 1.09 & 1.12             & 0.97          \\   
\multirow{4}{*}{$\tilde{\xi}_\mathrm{g+}$}    & \multirow{2}{*}{simple}    & $n_\star>300$                                & 1.05     & 1.07             & 0.98           \\  
                                                                        &                                                                    & $M_\star>10^{10.5}M_\odot/h $           & 1.11  & 1.1             & 1.0        \\  
                                                                        & \multirow{2}{*}{reduced}  & $n_\star>300$                                                                                & 1.13   & 1.12             & 1.01         \\  
                                                                        &                                                                         & $M_\star>10^{10.5}M_\odot/h $                                                                                 & 1.12  & 1.11             & 1.01         \\  \hline 
\end{tabular}
\begin{tablenotes}
\footnotesize
\item 
The mean gain in S/N when comparing the use of one versus three; one versus two or two versus three projections.
The gain is shown for $w_\mathrm{g+}$ and $\tilde{\xi}_\mathrm{g+,2}$, for simple and reduced shapes and for both shape samples, using the averaged covariance for two and three projections (compare to Table \ref{table:SNR gain}).
\end{tablenotes}
\label{table:SNR av cov}
\end{threeparttable}
\end{table}

\begin{figure}
    \centering
    \includegraphics[width=0.5\textwidth]{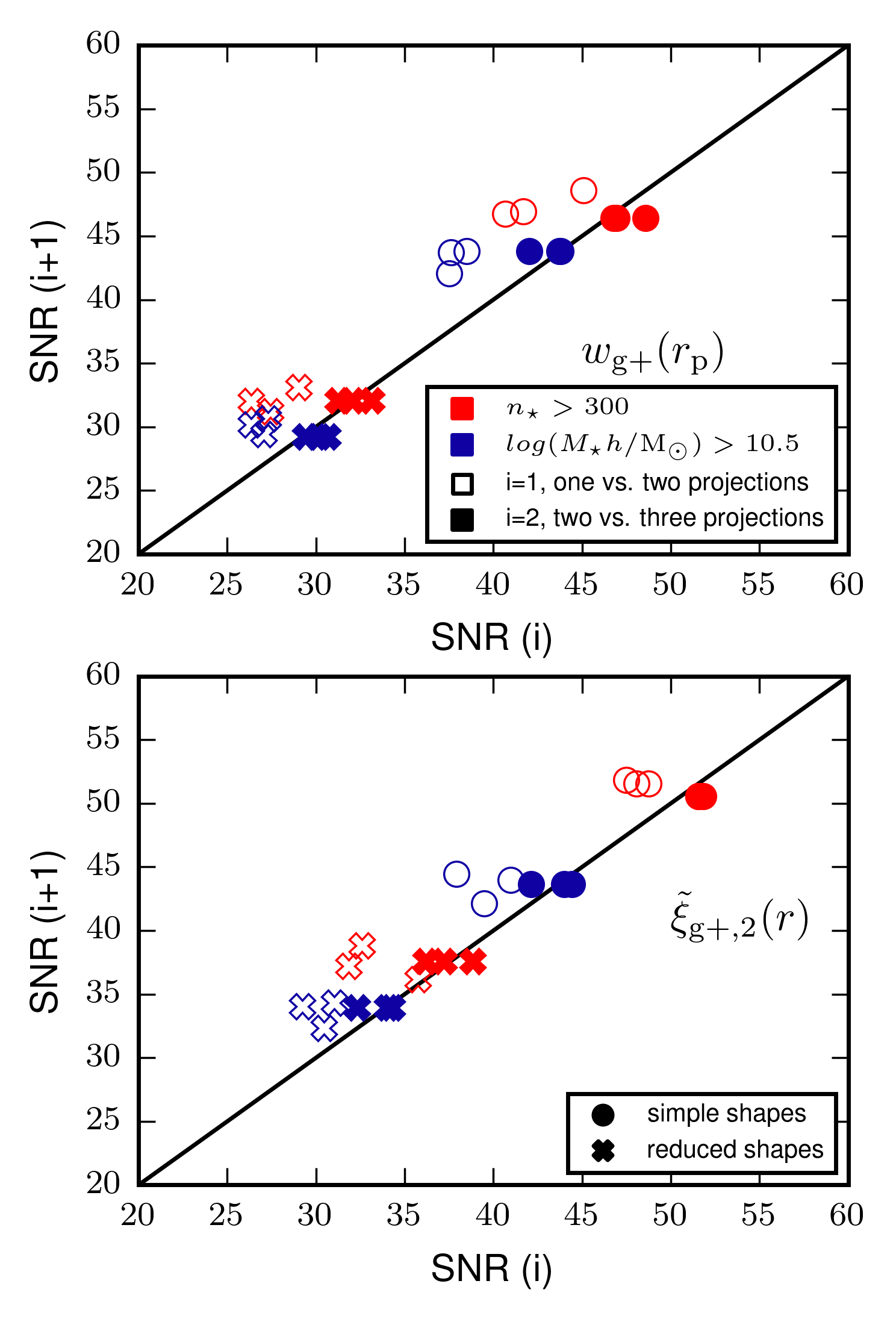}
    \caption{
    S/Ns for the $w_\mathrm{g+}$ (top) and $\tilde{\xi}_\mathrm{g+,2}$ (bottom) of combining one vs two (unfilled) and two vs three (filled) projections in TNG300 using the averaged covariance for two and three projections.
    The colours of the markers correspond to the different shape samples: $n_\star>300$ is red and $log(M_\star h/M_\odot)>10.5$ is blue.
    Shapes measured using the simple, reduced inertia tensor are depicted by circles, crosses, respectively.
    The black line denotes the boundary for which both S/Ns are equal.
    }
    \label{Fig:SNR av cov}
\end{figure}

Figure \ref{Fig:SNR av cov} shows the S/Ns of one versus two (unfilled markers) and two versus three (filled) for simple (circles) and reduced (crosses) shapes for $n_\star>300$ (red) and $log(M_\star h/M_\odot)>10.5$ (blue) for both $w_\mathrm{g+}$ (top) and $\tilde{\xi}_\mathrm{g+,2}$ (bottom).
Comparing Fig. \ref{Fig:SNR av cov} to Fig. \ref{Fig:SNR}, we see that the gains are slightly lower, but still above unity (black line) in most cases.
Table \ref{table:SNR av cov} presents the S/N gains in all cases mentioned previously.
Comparing Table \ref{table:SNR av cov} to Table \ref{table:SNR gain}, we see that the S/N gains are indeed slightly lower when averaging the covariance blocks, but they are still consistently above $1.0$ when increasing the number of projections from 1 to 3 or 1 to 2.
These gains are high enough to be considered robust, as even this lower bound scenario produces $>1$ S/N ratios.
Adding a third projection does not consistently add S/N, as we have already seen in other appendices.

\FloatBarrier

\FloatBarrier

\section{Fitted amplitude difference between multipole expanded and projected correlation function}
\label{app:fitted amplitude difference}

\begin{figure}
    \centering
    \includegraphics[width=0.5\textwidth]{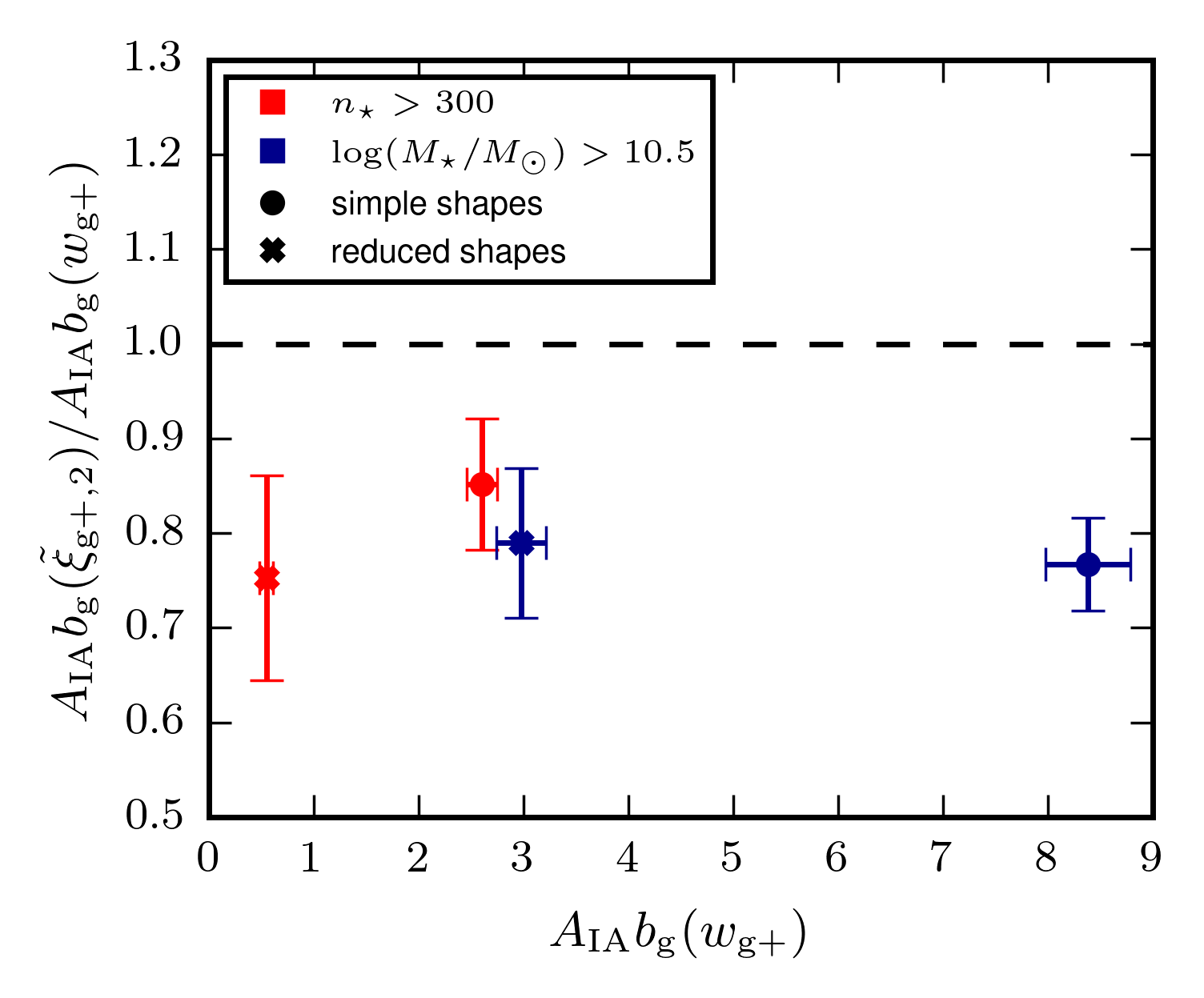}
    \caption{
    Fraction of $A_{IA}b_g$ obtained from $\tilde{\xi}_\mathrm{g+,2}$ over $A_{IA}b_g$ obtained from $w_\mathrm{g+}$ as a function of the $A_{IA}b_g$ obtained from $w_\mathrm{g+}$ using a $\Pi_\mathrm{max}=20\,\text{Mpc}/h$, depicted in red (blue) for the particle number (mass) cut, and as circles (crosses) for shapes obtained with the simple (reduced) intertia tensor. We show only the case of the joint amplitude across all three axes, since they are fully consistent with the amplitude of the individual axes.
    }
    \label{Fig:Relative Difference fitted amplitudes}
\end{figure}

\begin{figure}
    \centering
    \includegraphics[width=0.5\textwidth]{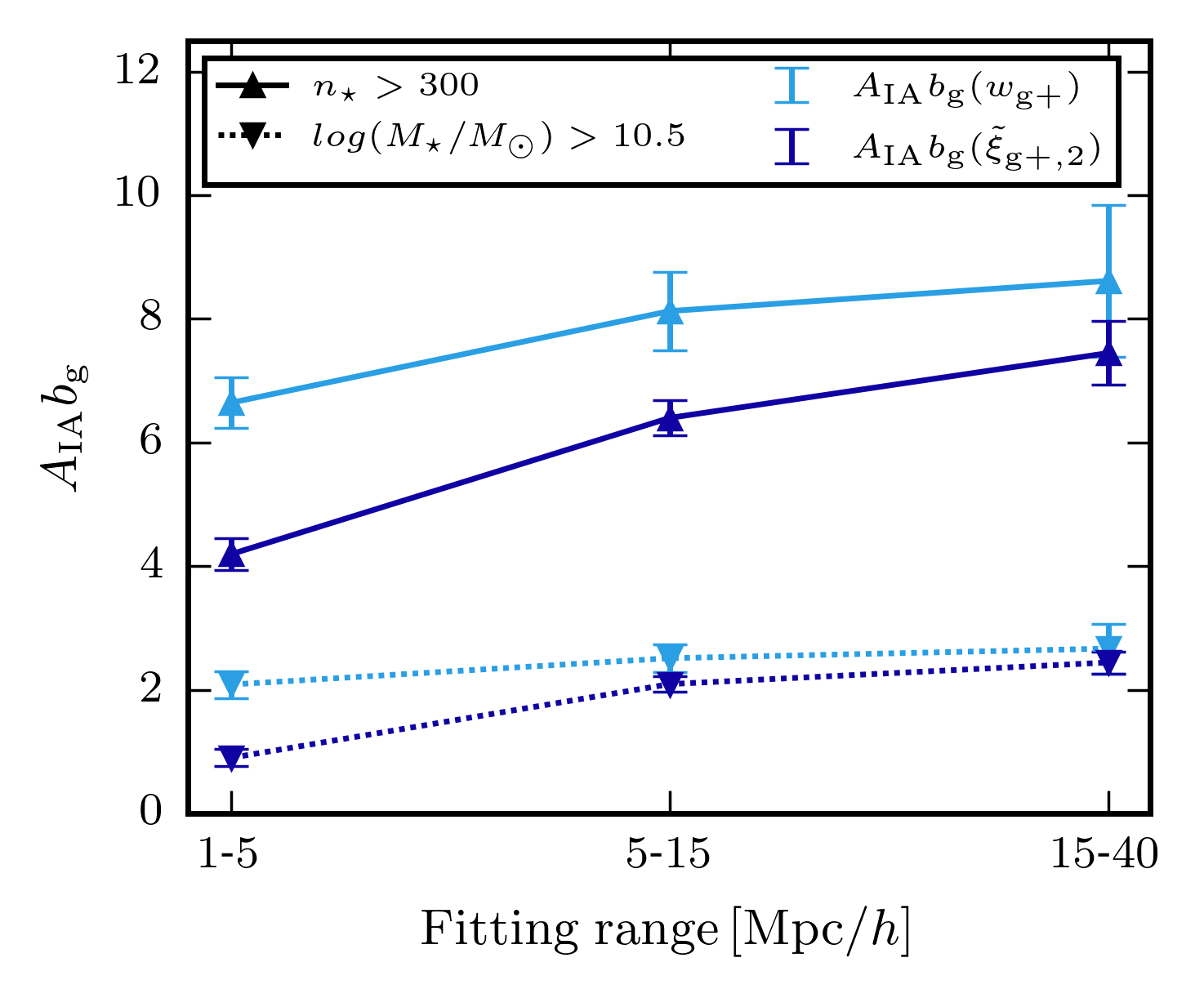}
    \caption{
    Evolution of NLA $A_{\rm IA}b_{\rm g}$ when fitting within different scale ranges, shown for particle number and mass cut with upward triangles and solid lines, and downward triangles and dashed lines, respectively. NLA amplitudes fitted with $\tilde{\xi}_\mathrm{g+,2}$ are depicted in dark blue and those fitted with $w_\mathrm{g+}$ in light blue, including the error bars.
    }
    \label{Fig:Fitting bias with scale}
\end{figure}

In this section, we investigate the difference between NLA amplitude obtained from multipole and projected correlation functions. In Fig. \ref{Fig:Relative Difference fitted amplitudes} we show the ratio of NLA amplitude between $\tilde{\xi}_\mathrm{g+,2}$ and $w_\mathrm{g+}$ as a function of $w_\mathrm{g+}$ NLA amplitude, for simple (reduced) inertia tensor shapes depicted as circles (crosses), and for the particle number (mass) sample cut in red (blue). The amplitudes from $\tilde{\xi}_\mathrm{g+,2}$ are consistently lower than those of projections. We believe that the most likely explanation for this is the difference in scale dependence. $\tilde{\xi}_\mathrm{g+,2}$ are functions of the full 3D distance magnitude, $r$, while $w_\mathrm{g+}$ are functions of the distance perpendicular to the line of sight $r_p\leq r$. The fitting range of $r_\mathrm{(p)}>6\,\mathrm{Mpc}/h$ was chosen close to the scale at which NLA is known to break down for $w_\mathrm{g+}$\,\citep{Tenneti_2015}. Hence, we can expect a model bias of $\tilde{\xi}_\mathrm{g+,2}$ if we fit the same scales. 
In addition,  we employed $\Pi_\mathrm{max}=20\,\mathrm{Mpc}/h$ for the main body of the paper, staying close to mildly non-linear scales in line-of-sight direction for $w_\mathrm{g+}$ with the advantage of decreasing shape noise at the same time.
However, such constraints are naturally not put on $\tilde{\xi}_\mathrm{g+,2}$.
Therefore, we would expect more similar signals for larger $\Pi_\mathrm{max}$ on large $r_\mathrm{(p)}$ since $w_\mathrm{g+}$ and $\tilde{\xi}_\mathrm{g+,2}$ then start probing similar, linear volumes.
To illustrate this effect, we additionally measure the correlations in TNG300 using $\Pi_\mathrm{max}=102.5\,\text{Mpc}/h$ up to scales of $r_\mathrm{max}=40\,\text{Mpc}/h$. Note that this increases the uncertainty on $w_\mathrm{g+}$, making this an unfit measurement for the main analysis of this work. Nevertheless, we can investigate the influence of a model mis-match by fitting the NLA amplitude in three different scale ranges: 1--5\,Mpc$/h$, 5--15\,Mpc$/h$, and 15--40\,Mpc$/h$. 

In Fig. \ref{Fig:Fitting bias with scale} we show the evolution of the fitted NLA amplitude for ranges of increasingly larger scales for $w_\mathrm{g+}$ and $\tilde{\xi}_\mathrm{g+,2}$ in light and dark blue, respectively, as well as for particle number cut sample with upward triangle symbols and solid line, and mass cut sample with downward triangle symbol and dashed line. We notice that the NLA fit to multipole expanded correlation functions has, as expected, smaller error bars than the fit to projected correlation functions, but also shows a larger mis-match on small scales as opposed to large scales where the NLA model is known to be reasonable. Correspondingly, for ranges of 15--40\,Mpc$/h$ the NLA amplitude is consistent between $\tilde{\xi}_\mathrm{g+,2}$ and $w_\mathrm{g+}$. This leads us to two conclusions: First, naively transferring the accumulated knowledge from modelling projected correlation functions to modelling multipole expanded correlation function will lead to biased results due to different scale dependent behaviour, and second, the bridge between large-scale models such as NLA and small scale models such as the halo model becomes ever more important when dealing with measurements where intermediate scales carry significant part of information, whether that is in simulation boxes of limited sizes or in real-world applications that increasingly rely on pushing scale cuts for improvement in cosmology inference.

\end{appendix}

\end{document}